\documentclass[10pt,journal]{IEEEtran} 

\usepackage[linesnumbered,ruled,vlined]{algorithm2e}
\usepackage{mathtools, cases}
\usepackage{textcomp}
\usepackage{makecell}
\usepackage{amsthm, enumitem}
\usepackage{amsmath,amsfonts,amssymb}
\usepackage{cite, url}
\usepackage{verbatim}
\usepackage{bm}
\usepackage{graphicx}
\usepackage{xcolor}
\usepackage{subcaption}
\usepackage{epstopdf}
\usepackage{mathrsfs}
\usepackage{txfonts}
\usepackage{caption}
\usepackage{lineno}
\usepackage{multirow}
\usepackage{booktabs}
\usepackage{color}
\usepackage{multicol}
\usepackage{stfloats}
\usepackage{diagbox}
\usepackage{esint}
\usepackage{float}
\usepackage{algpseudocode}
\usepackage{empheq}
\usepackage{tabularx}
\usepackage[a-1b]{pdfx}
\usepackage{flushend}

\allowdisplaybreaks[4]

\SetKwInput{KwReq}{\textbf{Input}}

\title{Communicate Less, Synthesize the Rest: Latency-aware Intent-based Generative Semantic Multicasting with Diffusion Models}

\author{Xinkai Liu, \textit{Student Member, IEEE}, Mahdi Boloursaz Mashhadi, \textit{Senior Member, IEEE}, Li Qiao, \textit{Member, IEEE}, Yi Ma, \textit{Senior Member, IEEE}, Rahim Tafazolli, \textit{Fellow, IEEE}, Mehdi Bennis, \textit{Fellow, IEEE}
\thanks{
Xinkai Liu, Mahdi Boloursaz Mashhadi, Li Qiao, Yi Ma, and Rahim Tafazolli are with 5GIC \& 6GIC, Institute for Communication Systems (ICS), University of Surrey, Guildford, United Kingdom (email: \{xinkai.liu, l.qiao, m.boloursazmashhadi, y.ma, r.tafazolli\}@surrey.ac.uk). Li Qiao is with the School of Information
and Electronics, Beijing Institute of Technology, Beijing 100081, China (e-mails: qiaoli@bit.edu.cn). Mehdi Bennis
is with the Centre for Wireless Communications, University of Oulu, 90014
Oulu, Finland (e-mail: mehdi.bennis@oulu.fi).

{A real-time on-device implementation of generative semantic multicasting with GAN-based synthesis was demonstrated at the IEEE ICCC 2025 \cite{ICCC}.}
}}

\begin{document}

\maketitle

\begin{abstract}  
      Generative diffusion models (GDMs) have recently shown great success in synthesizing multimedia signals with high perceptual quality enabling highly efficient semantic communications in future wireless networks. In this paper, we develop an intent-aware generative semantic multicasting framework utilizing pre-trained diffusion models. In the proposed framework, the transmitter decomposes the source signal to multiple semantic classes based on the multi-user intent, i.e. each user is assumed to be interested in details of only a subset of the semantic classes. To better utilize the wireless resources, the transmitter sends to each user only its intended classes, and multicasts a highly compressed semantic map to all users over shared wireless resources that allows them to locally synthesize the other classes, i.e. non-intended classes, utilizing pre-trained diffusion models. The signal retrieved at each user is thereby partially reconstructed and partially synthesized utilizing the received semantic map. We design a communication/computation-aware scheme for per-class adaptation of the communication parameters, such as the transmission power and compression rate to minimize the total latency of retrieving signals at multiple receivers, tailored to the prevailing channel conditions as well as the users’ reconstruction/synthesis distortion/perception requirements. The simulation results demonstrate significantly reduced per-user latency compared with non-generative and intent-unaware multicasting benchmarks while maintaining high perceptual quality of the signals retrieved at the users. {\color{black}For a typical setup of multicasting street scene images to 10 users, our proposed framework achieves a $15.4\%$ reduction in per-user latency at a fixed power budget, or equivalently $50\%$ reduction in the transmission power required to achieve a fixed per-user latency, compared with non-generative multicasting.}
\end{abstract}

\section{Introduction}\label{sec:intro}

{\color{black}\textit{Semantic Communication (Semcom)} is envisioned to play a crucial role in future wireless networks, specifically in emerging applications where communication of large multimodal signals with stringent latency and reliability constraints is required to reduce the wireless/network resources used, e.g. the wireless metaverse and digital twins, extended/mixed reality (XR/MR), and holographic teleportation and the internet of senses. In many such applications, different receivers may be interested in different parts of the signal according to their communication intent. Considering the limited wireless/network resources, the traditional multicasting schemes that transmit the same content to all users irrespective of their intent, perform inefficiently in such scenarios, thereby necessitating development of new intent-aware SemCom schemes for multicasting with intent heterogeneity. As an example, consider an XR/MR streaming scenario where the content needs to be simultaneously distributed to multiple users with different intents, or in the wireless metaverse, intent-aware multicasting is required to communicate the intended updates, states, or events of the virtual world to multiple users simultaneously in real-time to ensure all users remain synchronized. In many similar applications, multicasting to multiple users with different semantic interests is essential. Despite this need, the existing works in SemCom focus mostly on single-user point-to-point scenarios, and semantic multicasting remains less studied. In this work, we propose \textit{generative semantic multicasting}, where intent-aware decomposition of the source signal is carried out at the transmitter to allow communicating to each user only its intended part of the signal and synthesizing the rest locally with diffusion models, thereby avoiding unnecessary use of the wireless/network resources for transmission of the non-intended portions of the signal.}

Recently, the Generative AI (GenAI) models have shown great success in developing efficient low bitrate \textit{Generative Semantic Communication (Gen Semcom)} systems \cite{Li2024generative, xia2023generative, AR, Ece}. Generative models are capable of learning the general distribution of natural signals during training and synthesizing new samples with high perceptual quality at inference time. The generation process can be guided by multi-modal prompts and conditioning signals to produce high quality outputs with a desired semantic content. The emerging \textit{Generative Foundation Models} and \textit{Multimodal Large Language Models (MLLMs)}, e.g. Sora, Lumiere, and DALL.E, etc., provide ample opportunities to develop efficient generative Semcom frameworks \cite{TokCom, xu2024semantic, Foundation1, Large, LightweightLAI}. As these models possess a vast general knowledge captured via intensive pre-training on huge amount of data, they alleviate the need for a shared Knowledge Base (KB) between the semantic transmitter and receiver, thereby reducing the corresponding knowledge sharing overheads. This vast general knowledge also makes the generative Semcom framework applicable to various datasets and tasks thereby achieving universality. The GenAI models are trained to maximize the perceptual quality of the synthesized signal, and the theoretical limits of generative Semcom are governed by the \textit{rate-distortion-perception} theory \cite{RDP, RDP2}. This theory explores the threefold trade-off between rate, distortion, and the perceptual quality of the synthesized signal. While the rate-distortion-perception function is analytically derived only for a few source distributions, it is generally estimated empirically for natural signals, e.g. images, audio/video, point cloud, etc., by training deep source encoders with a perceptual loss function, e.g. Wasserstein distance, on large datasets. In generative Semcom, the transmitter extracts the intended semantics, e.g. in form of textual prompts \cite{Prompt1, Prompt2, Genvideosem}, compressed embeddings \cite{Giordano2024generative, yilmaz2024high}, semantic map \cite{LightDiff, WU2024519}, edge map \cite{Li2024generative, xu2024semantic}, etc., which are then transmitted over the channel. The receiver uses these semantics to guide a generative model, synthesizing a signal that is semantically consistent and highly realistic. {\color{black}Despite the above works, multiuser generative semantic communications, specifically in the multicasting setup, and its scalability to increased number of users, considering the communication/computation latency remains less studied.}

\begin{table*}[t]
         \centering
         \begin{center}
         \normalsize
        \begin{tabular}{|l|l|l|l|l|l|l|}
        \hline
        \textbf{ } &
        \color{black}\textbf{\cite{R1}}&
        \color{black}\textbf{\cite{R2}}&
        \color{black}\textbf{\cite{R3}}&
        \color{black}\textbf{\cite{R4}}&
        \color{black}\textbf{This Work}\\
        \hline\hline
        \color{black}\textbf{JSCC or SSCC?} & \color{black}SSCC & \color{black}JSCC & \color{black}SSCC & \color{black}JSCC & \color{black}SSCC \\\hline 

        \color{black}\textbf{E2E Training Needed?} & \color{black}Yes & \color{black}Yes & \color{black}Yes & \color{black}Yes & \color{black}No (Pre-Training) \\\hline

        \color{black}\textbf{Gen or Non-Gen?} & \color{black}Non-Gen & \color{black}Non-Gen & \color{black}Non-Gen & \color{black}Gen (MAE) & \color{black}Gen (Diffusion) \\\hline
        
        \color{black}\textbf{Intent/Task Heterogeneity?} & \color{black}Yes & \color{black}Yes & \color{black}No & \color{black}Yes & \color{black}Yes \\\hline

        \color{black}\textbf{Comm/Comp. Latency Aware?} & \color{black}No & \color{black}No & \color{black}No & \color{black}No & \color{black}Yes \\\hline
        \end{tabular}
        \end{center}
         
         \caption{\color{black}Key Differences Between this Work and Semantic Multicasting/Broadcasting Literature.}
         \label{tab:Lit}
     \end{table*}

In this paper, \textit{we develop an scalable latency-aware generative semantic multicasting framework with adaptive intent-based resource allocation utilizing pre-trained diffusion models.} We claim three key benefits for the proposed framework: Firstly, generative multicasting in our Semcom framework improves utilization of the wireless and network resources compared with non-generative and intent-unaware multicasting benchmarks. This also reduces the total latency while maintaining the reconstruction/synthesis distortion/perception quality. Secondly, the adoption of pre-trained models allows a source-channel separation-based Semcom architecture, thereby alleviating the need for end-to-end joint training of the transmitter and receiver, which is required in many existing Semcom frameworks \cite{SemCom1, Semcom2, pointcloudsem, SemCom4}. Such a separation-based architecture offers improved adaptability to varying channel conditions, improved scalability to increase the number of users, and better compatibility with the existing design of wireless communication networks. {\color{black}Finally, diffusion models enable synthesizing photorealistic outputs with better diversity, realism, and fine-grained details. TABLE I provides an overview of the major differences between this work and the key existing works on semantic multicasting/broadcasting \cite{R1, R2, R3, R4}. To the best of our knowledge, this work is the first of its kind to design joint communication/computation latency-aware intent-based generative semantic multicasting with pre-trained diffusion models, based on source-channel coding separation to achieve scalable and adaptable semantic multicasting compatible with the existing wireless networks, and derive the corresponding rate-distortion/perception tradeoffs.} {\color{black}The contributions of this work are thereby three-fold:
    \begin{itemize}
        \item To achieve intent-awareness, we propose a semantic decomposition scheme which splits the source signal into multiple sub-signal classes at the transmitter based on the multi-user intent feedback, i.e. each user is assumed to need only a subset of the semantic classes based on its communication intent. We develop a mixed reconstruction/synthesis scheme with semantic diffusion models, where users partially reconstruct and partially synthesize the signal from a highly compressed semantic map, i.e. the intended sub-signal classes are reconstructed and the non-intended sub-signals classes are locally synthesized by the pre-trained generative diffusion model. The transmitter broadcasts the semantic map to all users over shared wireless resources, thereby utilizing orthogonal resources only to transmit the sub-signal classes intended for each user, which further saves the wireless resources.
        
        \item To achieve scalability and adaptability, different from majority prior SemCom research, we follow a septate source-channel coding scheme and pre-train a generative diffusion model with classifier-free semantic guidance. Accordingly, we derive the achievable rate-distortion/perception curves for our proposed mixed reconstruction/synthesis scheme. We then use these curves to determine and adapt the communication parameters, i.e. the transmission power and compression rate in our proposed framework on the go, without requiring end-to-end retraining. We show efficient scalability and adaptability of our proposed framework with extensive simulations.
        
        \item To achieve latency-awareness, we design a scheme for per-class adaptation of the communication parameters to minimize the total latency considering both the communication and computation time. As the signal retrieved at each user is partially reconstructed and partially synthesized, we formulate optimization of the communication parameters for latency-aware synchronisation of the synthesized and reconstructed datastreams, based on the prevailing channel conditions, users' intents and their reconstruction/synthesis distortion/perception requirements. We propose a computationally efficient sequential quadratic programming (SQP)-based algorithm for solving the resulting optimization problem, demonstrating s a 15.4\% reduction in per-user latency at a fixed power budget, or equivalently 50\% reduction in the transmission power required to achieve a fixed per-user latency compared with intent-unaware and non-generative multicasting, for a typical image semantic multicasting setup of street scene images with 10 users.
    \end{itemize}}

The rest of this paper is organized as follows. In Section \ref{SEC_Proposed}, we present our proposed diffusion-based generative semantic communication framework and the corresponding intent-based latency-aware adaptive resource allocation scheme. Simulation results are provided in Section \ref{SEC_Result}, and we conclude the paper in Section \ref{SEC_Conclusion}.

\textit{\textbf{Notations}}: Boldface lower and upper-case symbols denote column vectors and matrices, respectively. Calligraphic letters denote mathematical operators. $[K]$ is $\{1,2,...,K\}$, and $\left| {\bf v} \right|$ denotes the length of vector ${\bf v}$. Additionally, $\oplus$ and $\odot$ denote point/pixel/voxel-wise summation and product, and $\bigcup$ is the union operator. Expectation is denoted by $\mathbb{E}[.]$. Finally, $\mathcal{U}[.,.]$ and $\mathcal{N}[.,.]$ denote the uniform and Gaussian probability distributions, respectively.

\begin{figure*}[tb]
\centerline{\includegraphics[scale=0.32]{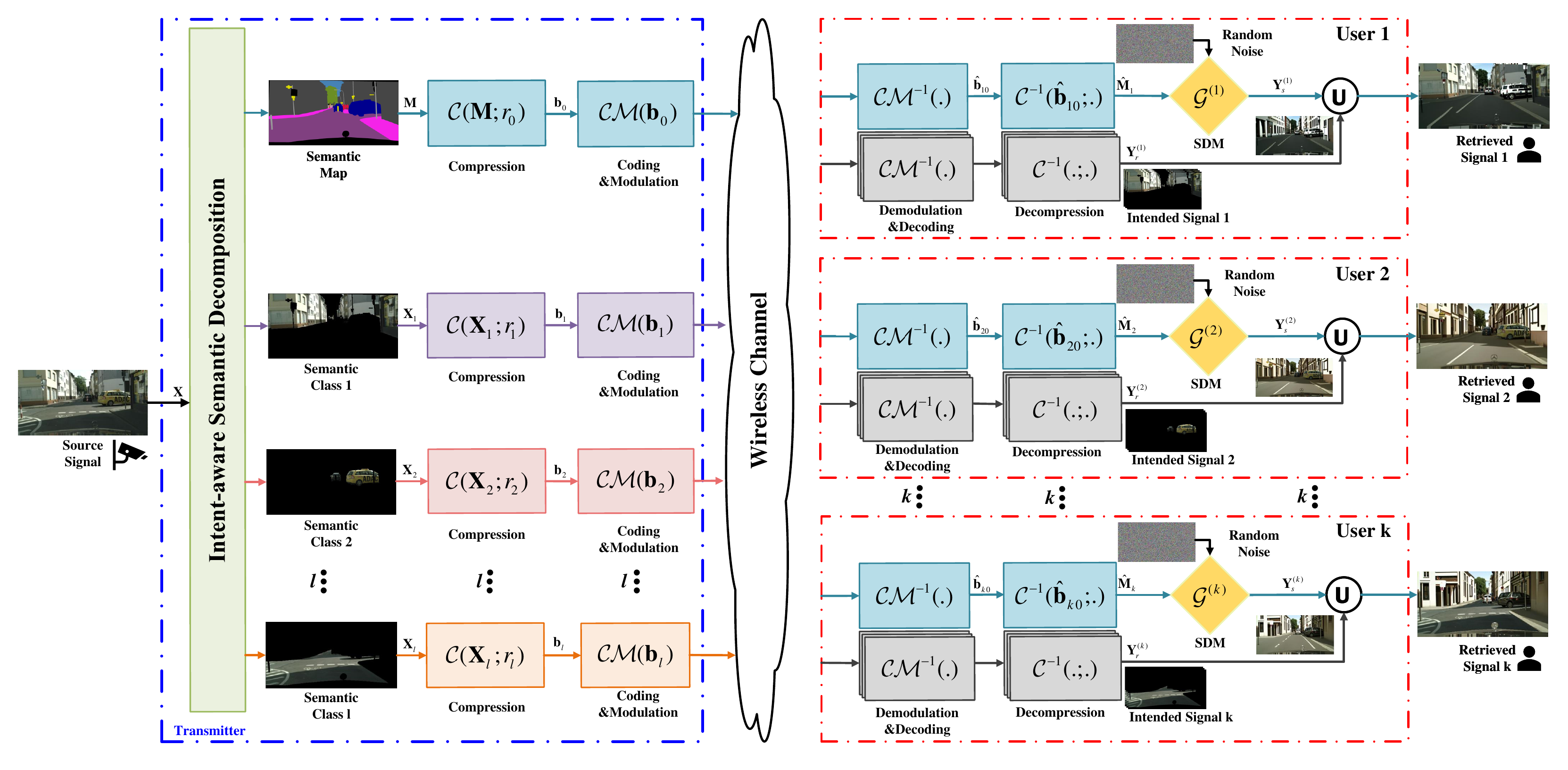}}
 \captionsetup{font={footnotesize}, singlelinecheck = off, name={Fig.},labelsep=period}
\caption{Proposed Framework for Diffusion-based Generative Semantic Multicasting with Intent-aware Semantic Decomposition.}
\label{fig_proposed01}
\end{figure*}

\section{Intent-based Generative Semantic Multicasting} \label{SEC_Proposed}
We consider a transmitter serving $K$ users in a multicasting setup as depicted in Fig. \ref{fig_proposed01}. The source signal to be transmitted to the users contains $L$ semantic classes. For example, assuming the source signal is images captured by a street scene camera, the semantic classes include ``car", ``person", ``building", ``sky", etc. Denoting the source signal by ${\bf X}$, we have ${\bf X}=\bigcup_{l=1}^{L} {\bf X}_l$, where ${\bf X}_l$ represents the $l$'th class. Each user is only interested in a specific subset of the classes based on its goal or intent. For example, a user carrying out ``traffic surveillance" based on the received signal is interested in the ``car" class but not ``sky". This user may need details of the car class, e.g. the cars' make and model, as well as their color and number plate. A user carrying out ``weather monitoring" is only interested in details of the ``sky" class, and another user carrying out ``Path Planning" for Intelligent Transportation Systems (ITS) will be interested in the ``street signs" and the ``road" classes. Hence, we define the multi-user intent matrix ${\bf I}_{K \times L}$, where $[{\bf I}]_{kl}=1$ if user $k$ is interested in class $l$, otherwise $[{\bf I}]_{kl}=0$. The intended portion of the source signal for use $k$, i.e. the semantic classes each user is interested in, is thereby written as ${\bf X}^{(k)}_I = \{\bigcup_l {\bf X}_l | [{\bf I}]_{kl}=1\}$. Similarly, the non-intended portion of the source signal for use $k$ is thereby written as ${\bf X}^{(k)}_N = \{\bigcup_l {\bf X}_l | [{\bf I}]_{kl}=0\}$.

Fig. \ref{fig_proposed01} depicts our proposed framework for intent-aware generative semantic multicasting with pre-trained diffusion models. This framework includes multi-class semantic decomposition for mixed reconstruction/synthesis with denoising diffusion models, per-class adaptive intent-aware multicasting, and communication/computation-aware semantic latency minimization with multi-stream synchronization, as explained in the following subsections.

\subsection{Multi-class Semantic Decomposition for Mixed Reconstruction/Synthesis with Semantic Diffusion Models}\label{SDMGDM}
\subsubsection{Multi-class Semantic Decomposition with DDRNets}
\begin{figure*}[t]
     \centerline{\includegraphics[scale=0.33]{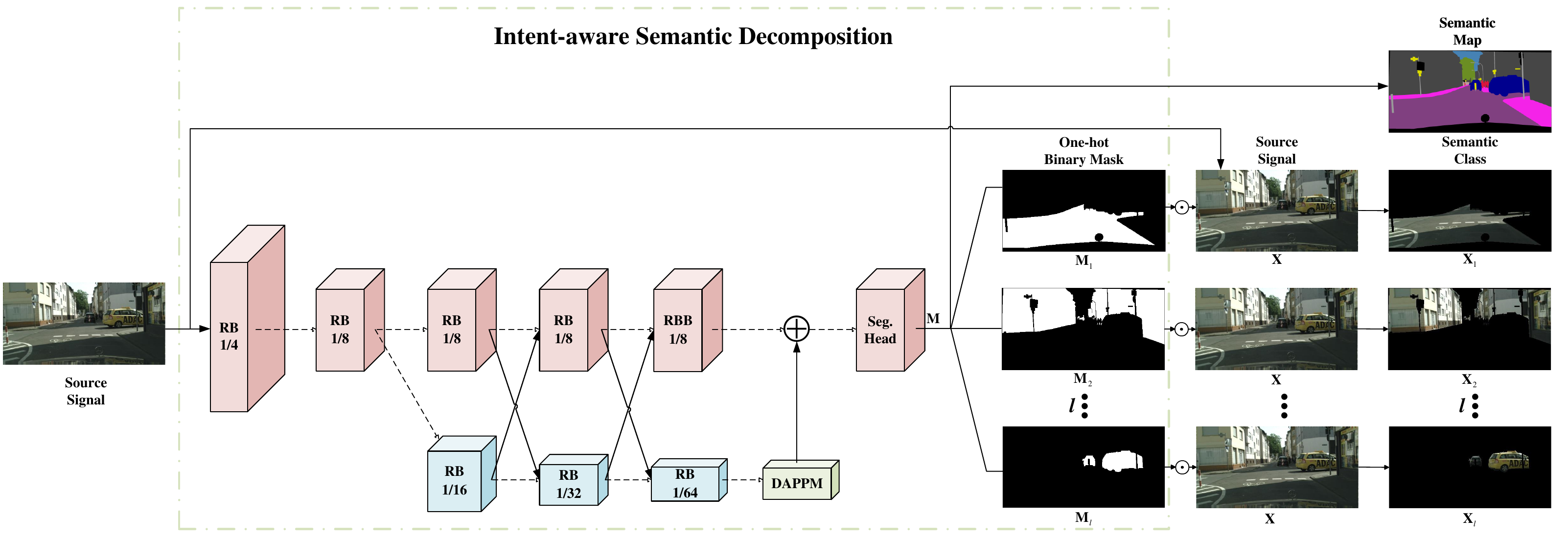}}
      \captionsetup{font={footnotesize}, singlelinecheck = off, name={Fig.},labelsep=period}
     \caption{\color{black}Intent-aware semantic Decomposition with DDRNets for generative multicasting.}
     \label{fig:DDRNets}
      
\end{figure*}

At the transmitter, a semantic segmentation model is used to generate a semantic map ${\bf M}$ from the source signal as depicted in Fig. \ref{fig_proposed01} for sample image signals. To do this, any existing semantic segmentation model, e.g. \cite{Seg2, Seg1, SAM1}, can be utilized, and the choice of the appropriate model depends on the required accuracy and affordable computational complexity. {\color{black}To minimize the end-to-end latency, we use a deep dual-resolution network (DDRNet)\cite{Seg1}-based architecture. DDRNets are efficient architectures that were proposed to achieve real-time performance for semantic segmentation of the street scenes. DDRNets are composed of two deep convolutional branches, and a deep information extractor called Deep aggregation pyramid pooling module (DAPPM) to enlarge effective receptive fields and fuse multi-scale context based on low-resolution feature maps as shown in Fig. \ref{fig:DDRNets}. RB and RBB represent sequential and single residual bottleneck blocks. Information is extracted from low-resolution feature maps e.g., $1/8$ denotes high-resolution branch create feature maps whose resolution is 1/8 of the input image resolution. The black solid lines demonstrate the data processing path (including up-sampling and down-sampling) and black dashed lines denote information paths without data processing. The architecture is trained with the cross entropy loss to classify each image pixel into a semantic class, thereby generating the semantic map ${\bf M}$.} The semantic map is then converted to multiple one-hot binary masks ${\bf M}_l$. With this representation, we have ${\bf X}_l={\bf M}_l \odot {\bf X}$. The intended mask of user $k$ is then ${\bf M}^{(k)}_I = \{\bigcup_l {\bf M}_l | [{\bf I}]_{kl}=1\}$ and we have ${\bf X}^{(k)}_I={\bf M}^{(k)}_I \odot {\bf X}$. Similarly, the non-intended  mask of user $k$ is then ${\bf M}^{(k)}_N = \{\bigcup_l {\bf M}_l | [{\bf I}]_{kl}=0\}$ and we have ${\bf X}^{(k)}_N={\bf M}^{(k)}_N \odot {\bf X}$. Note that knowledge of the intent matrix ${\bf I}$ is assumed at the transmitter. 
Our proposed framework takes a mixed approach by using a conventional lossy compression scheme for the intended portion of the source, and generative AI for the other semantic classes, i.e. the non-intended portion, at the receiver. Hence, the signal retrieved at each user is partially reconstructed and partially synthesized with generative AI based on the multi-user intent. This allows to maintain a high fidelity for the intended classes while significantly reducing the communication overhead by transmitting the other classes over shared wireless resource at a low rate with a generative scheme. 


\subsubsection{Mixed Reconstruction/Synthesis with Semantic Diffusion Models}
At the receiver, each user receives two datastreams, one containing its intended classes which is decompressed to reconstruct its intended portion of the source signal, and the other containing the semantic map ${\bf M}$ which is fed into a pre-trained generative diffusion model for high-fidelity synthesis of its non-intended portion. The signal retrieved at user $k$ is thereby written as ${\bf Y}^{(k)}={\bf Y}^{(k)}_r \bigcup {\bf Y}^{(k)}_s$, where ${\bf Y}^{(k)}_r$ is the reconstructed and ${\bf Y}^{(k)}_s$ is the synthesized portion. {\color{black}Any state-of-the-art generative models with semantic guidance can be used to locally synthesize the non-intended signal portions ${\bf Y}^{(k)}_s$ at the users. Here we pre-train GDMs with classifier-free guidance for high quality/fidelity image synthesis from the semantic map at each user. {\color{black} Note that we adopt diffusion models in the proposed framework as they enable synthesizing photorealistic outputs with better diversity, realism, and fine-grained details, thereby significantly improving users' quality of experience, e.g. in comparison with other classes of generative models, e.g. GANs, VAEs, etc. We provide comparisons with GAN-based synthesis of the non-intended classes later in simulation results. Refer to \cite{CDDM, Dawn, Reinforcement, Matti, SurveyDiff} for details on diffusion models and their various recent applications in wireless communications.}



    Dropping the user index $k$ for brevity, the forward (diffusion) process is defined via 
\begin{align}
    \rho^{\mathrm{Diff}}({\bf Y}_t \mid {\bf Y}_{t-1}) = \mathcal{N}({\bf Y}_t ; \sqrt{1 - \beta_t} {\bf Y}_{t-1}, \beta_t I),
\end{align}
where ${\bf Y}_t$ is the signal at the $t$'th diffusion step, $\rho^{\mathrm{Diff}}({\bf Y}_t \mid {\bf Y}_{t-1})$ is the diffusion distribution, and $\beta_t \in (0, 1)$ is the trainable variance schedule controlling the amount of noise added at each step. To synthesize ${\bf Y}_s={\bf Y}_0$, the reverse (synthesis) process learns to remove the noise iteratively, transforming the noisy input ${\bf Y}_{\tau}$ to the synthesized image ${\bf Y}_0$, using the additional conditioning information from the received semantic map ${\bf \hat{M}}$. {\color{black}The synthesis process is modeled by a parameterized Markov chain
\begin{align}
    \rho^{\mathrm{Syn}}_\theta({\bf Y}_{t-1} \mid {\bf Y}_t, {\bf \hat{M}}) &= \mathcal{N}({\bf Y}_{t-1} ; \mu_\theta({\bf Y}_t, {\bf \hat{M}}, t), \Sigma_\theta({\bf Y}_t, {\bf \hat{M}}, t)), \\ \nonumber
    \rho^{\mathrm{Syn}}_\theta({\bf Y}_{0:\tau} \mid {\bf \hat{M}}) &= p({\bf Y}_{\tau}) \prod_{t=1}^{\tau} p_\theta({\bf Y}_{t-1} \mid {\bf Y}_t, {\bf \hat{M}}),
\end{align}
where $\rho^{\mathrm{Syn}}_\theta({\bf Y}_{t-1} \mid {\bf Y}_t, {\bf \hat{M}})$ is a Gaussian distribution with learned parameters $\mu_\theta({\bf Y}_t, {\bf \hat{M}}, t)$ and $\Sigma_\theta({\bf Y}_t, {\bf \hat{M}}, t)$, in which $\mu_\theta$ and $\Sigma_\theta$ are learned functions that depend on the noisy image ${\bf Y}_t$, the semantic map ${\bf \hat{M}}$, and the current timestep $t$. To learn these parameters, we use a U-Net architecture where the noisy image is passed through the encoder, consisting of a series of downsampling ResBlocks and attention layers. These blocks operate at various resolutions to capture spatial features at multiple scales. The semantic map information is injected into the decoder using SPADE architectures \cite{8953676}. At each layer of the decoder, SPADE scales the features according to the semantic map, embedding semantic information in a spatially adaptive manner. This allows the network to maintain a strong correlation between the generated image and the input semantic map. The goal of training is to minimize the variational lower bound (VLB) of the negative log-likelihood. This can be broken down into a denoising objective by optimizing a noise prediction model. Specifically, the model learns to predict the noise \( \epsilon \) at each step, and the loss is formulated as
\begin{align}
    \mathbb{E}_{t, {\bf Y}_0, \epsilon} \left[ \lVert \epsilon - \epsilon_\theta (\alpha_t {\bf Y}_0 + \sqrt{1 - \alpha_t} \epsilon, {\bf \hat{M}}, t) \rVert^2 \right],
\end{align}
where \( \epsilon \) is the true noise sampled from a Gaussian distribution, \( \epsilon_\theta \) is the noise predicted by the model at time step \( t \), given the noisy image \( {\bf Y}_t \), the semantic map ${\bf \hat{M}}$, and the time step \( t \).} During inference, the diffusion process iteratively denoises the noisy image input \( {\bf Y}_{\tau} \) to generate the final image \( {\bf Y}_0 \). At each step \( t \), the model predicts the noise \( \epsilon_\theta({\bf Y}_t, {\bf \hat{M}}, t) \) and updates the image as
\begin{align}
   {\bf Y}_{t-1} = \frac{1}{\sqrt{1 - \beta_t}} \left( {\bf Y}_t - \frac{\beta_t}{\sqrt{1 - \alpha_t}} \epsilon_\theta({\bf Y}_t, {\bf \hat{M}}, t) \right) + \sigma_t {\bf N},
\end{align}

where \( {\bf N} \sim \mathcal{N}(0, I) \) is Gaussian noise, and \( \sigma_t \) is a trainable parameter controlling the noise scale at each step. For more information on diffusion models and semantic guidance refer
to \cite{SDM2025}.


    

\subsection{Intent-Aware Generative Multicasting}
To use the wireless resources efficiently, the transmitter multicasts the semantic map ${\bf M}$ to all the users over shared wireless resources. This map will be utilized at the users to locally synthesize the non-intended signal portions, i.e. ${\bf Y}^{(k)}_s$, using denoising diffusion models. With this idea, the transmitter only needs to send the intended portions $\{{\bf X}^{(k)}_I\}_{k=1}^{K}$ over orthogonal channels, thereby saving the wireless resources. The semantic map and the user-intended signals $\{{\bf X}^{(k)}_I\}_{k=1}^{K}$ are separately coded, modulated, and transmitted over different channels to enable rate and power adaptation based on the multi-user intent.

We assume Rayleigh block fading where the channel gain of the $k$-th user is expressed as $h_{k} = \sqrt{\epsilon_o \left(d_k \right)^{-\varphi}} \tilde{h}_k,~\forall k\in [K],$ where $\tilde{h}_k$ is a random scattering element captured by zero-mean and unit-variance circularly symmetric complex Gaussian (CSCG) variables, $\epsilon_o$ is the path loss at the reference distance $d =1$ m, $\varphi$ is the path loss exponent, and $d_k$ is the distance of user $k$ from the transmitter. 




\subsubsection{Multicasting the Semantic Map over Shared Wireless Resources} The semantic map is compressed by ${\bf b}_0=\mathcal{C}({\bf M}; r_0)$, where $\mathcal{C}(.; r_0)$ denotes the compression operator, $r_0$ is an adaptable compression rate in bit per pixel (bpp), and ${\bf b}_0$ is the resulting datastream which is then multicasted to all the users over shared wireless resources.\footnote{Various compression techniques, e.g. JPEG\cite{JPEG}, BPG \cite{BPG}, NTC \cite{balle2020nonlinear}, etc., can be used based on the type of the source signal to be transmitted and design requirements.} The compression rate $r_0$ controls the resulting datasize of the semantic map, i.e. ${\left| {{{\bf{b}}_0}} \right|} = r_0 \times {\left| {\bf X} \right|}$, and is adapted based on the wireless channel conditions and the received signal-to-noise ratios (SNRs) of the users. The resulting datastream is then encoded and modulated over the channel, with $\mathcal{CM}(.)$ representing the modulation and coding operator. Denoting the transmit power of the semantic map by $p_0$, the resulting data rate of the semantic map at user $k$ is given by 
\begin{align}\label{eq:rate1}
    R_{k0}(p_0) = B_0 {\log _2}\left( 1 + \frac{ p_{0}|h_{k}|^2}{B_0 N_0} \right), k\in[K],
\end{align}
in which $\eta_{k0}(p_0) = \frac{ p_{0}|h_{k}|^2}{B_0 N_0},~\forall k\in [K]$ is the received SNR of the semantic map at user $k$, where $B_0$ denotes the multicast channel bandwidth and $N_0$ is the noise power spectral density. The communication latency of the semantic map at user $k$ is thereby ${T_{k0}}(p_0, r_0) = \frac{r_0 \times {\left| {\bf X} \right|}}{{{R_{k0}}(p_0)}}$, $\forall k\in[K]$. 

At each user, the signal received from the channel for the semantic map is decoded and demodulated, with $\mathcal{CM}^{-1}(.)$ representing the demodulation and decoding operator. Denoting the datastream of the semantic map received at user $k$ by ${\hat{\bf b}}_{k0}$, the semantic map recovered at user $k$ is given by ${\hat{\bf M}}_k=\mathcal{C}^{-1}({\hat{\bf b}}_{k0}; r_0)$, where $\mathcal{C}^{-1}(.; r_0)$ is the decompression operator. The recovered semantic map is then input to the semantic-guided generative model, and thereby the synthesized portion of the signal retrieved at the $k$'th user is given by ${\bf Y}^{(k)}_s={\bf M}^{(k)}_N \odot \mathcal{G}^{(k)}({\hat{\bf M}}_k)$, where $\mathcal{G}^{(k)}$ denotes the pre-trained generative model at user $k$.

\subsubsection{Orthogonal Transmission of the User Intended Classes} Each user-intended class is similarly compressed by ${\bf b}_l=\mathcal{C}({\bf X}_l; r_l)$ to get the corresponding datastream ${\bf b}_l$, which is then transmitted to users over orthogonal wireless channels\footnote{Hybrid (non/)orthogonal multiple access techniques, e.g. RSMA, for the (non/)intended subsignal classes can be adopted in future work to further improve the performance of our proposed generative multicasting framework.}, with $\mathcal{CM}(.)$ representing the modulation and coding operator.

The adaptive compression rate $r_l$ controls the resulting datasize of each user-intended class, i.e. ${\left| {{{\bf{b}}_l}} \right|} = r_l \times {\left| {\bf \Bar{X}} \right|}$, where ${\left| {\bf \Bar{X}} \right|}=\sum_{l=1}^{L}\left| {\bf X}_l \right| / L$ is the average number of pixels per class, and $\left| {\bf X}_l \right|$ is the average number of pixels of the $l$'th class. Denoting the transmit power of the $l$'th semantic class by $p_l$, the resulting data rate for user-intended class $l$ at user $k$ is 
\begin{align}\label{eq:rate2}
    R_{kl}(p_l) = B_l {\log _2}\left( 1 + \frac{ p_{l}|h_{k}|^2}{B_l N_0} \right), k\in [K], l\in[L],
\end{align}
in which $\eta_{kl}(p_l) = \frac{ p_{l}|h_{k}|^2}{B_l N_0},~\forall k\in [K], l\in[L]$ is the received SNR of intended class $l$ at user $k$, where $B_l$ denotes the channel bandwidth allocated to intended class $l$ and $N_0$ is the noise power spectral density. The average communication latency for intended class $l$ at user $k$'s is thereby ${T_{kl}}(p_l, r_l) = \frac{r_l \times {\left| {\bf \Bar{X}} \right|}}{{{R_{kl}}(p_l)}}$, $\forall k\in [K], l\in[L]$. We represent the demodulation and decoding operator by $\mathcal{CM}^{-1}(.)$. Finally, denoting the received datastream of intended class $l$ at user $k$ by ${\hat{\bf b}}_{kl}$, the corresponding intended portion reconstructed at user $k$ is given by ${\bf Y}^{(kl)}_r=\mathcal{C}^{-1}({\hat{\bf b}}_{kl}; r_l)$, and we have ${\bf Y}^{(k)}_r = \{\bigcup_l {\bf Y}^{(kl)}_r | [{\bf I}]_{kl}=1\}$.

\subsection{Adaptive Communication/Computation-aware Semantic Latency Minimization with Multi-Stream Synchronization}
In this subsection, we propose a communication/computation-aware approach for selecting the optimal communication parameters, such as the transmission power and compression rate to minimize the total latency of retrieving the signals at the users, tailored to the prevailing channel conditions as well as the users' reconstruction/synthesis distortion/perception requirements.

The total latency to retrieve user $k$'s signal is the maximum of the two latencies to retrieve its reconstructed and synthesised portions, each of which are the summation of the corresponding communication and computation latencies. The communication latencies, i.e. $T_{k0}, T_{k}$, depend on the corresponding transmission power and compression rate as calculated in the previous subsection. The computation latency is dominated by the latency due to running of the pre-trained generative model to synthesize the non-intended portion of the signal at each user from the received semantic map denoted by $T_k^g$. Hence, the computation latency depends on the number of Floating Point Operations (FLOPs) required to run the adopted generative model, and the computation power of the processor at each user in terms of FLOPs per Second. 

To retrieve the signals, multiple data streams corresponding to the semantic map and the intended classes should be synchronised at each user. This synchronisation should take into account both the communication and computation latency for multiple streams, and as the generation time of the diffusion model $T_k^g$ dominates the computation latency, it becomes specifically crucial. Thereby, the resulting latency to retrieve the signal at user $k$, considering multi-stream synchronisation is given by $T_k=\max\left\{T_k^g+{T_{k0}}(p_0,r_0), [{\bf I}]_{kl} T_{kl}(p_l, r_l), l\in[L]\right\}$.

The reconstruction and synthesis distortion/perception requirements are measured by $\Phi_r\left(r_l\right)$ and $\Phi_s\left(r_0\right)$ metrics, respectively. These can be any distortion or perception metrics, e.g. MS-SSIM \cite{SSIM}, LPIPS \cite{LPIPS}, FID \cite{FID}, etc. According to the rate-distortion-perception theorem \cite{RDP2}, $\Phi_r\left(r_l\right)$ and $\Phi_s\left(r_0\right)$ are decreasing functions of the compression rates, i.e. $r_l, r_0$. Minimization of the sum users latency with the intent matrix is thereby formulated as 
\begin{equation}\label{OptiProb}
\begin{aligned}
&\min_{\left\{p_0, r_0, p_{l}, r_{l}, l\in[L]\right\}} & & \sum_{k=1}^K \max\left\{T_k^g+{T_{k0}}(p_0,r_0), [{\bf I}]_{kl} T_{kl}(p_l, r_l), l\in[L]\right\} \\
& \text{s.t.} & &  \sum\nolimits_{l=0}^L p_{l}\leq P_{\text{T}},\\
&&& \Phi_r\left(r_l\right) \leq \min \left\{\varepsilon_{kl}^{(r)} \big{|} [{\bf I}]_{kl}\neq 0, k\in[K]\right\},~ \forall l\in[L],\\
&&& \Phi_s\left(r_0\right) \leq \min \left\{\varepsilon_k^{(s)}, k\in[K]\right\},
\end{aligned}
\end{equation}
where $P_{\text{T}}$ is the maximum transmit power budget, and $\varepsilon_k^{(r)}$ and $\varepsilon_k^{(s)}$ denote the distortion/perception requirements at user $k$, respectively.\footnote{It is assumed that a smaller value of the metric represents improved reconstruction/synthesis quality. Many distortion/perception metrics can be easily normalized and transformed to satisfy this.} {\color{black}To solve optimization (\ref{OptiProb}), we introduce the auxiliary variables $\{z_k\}_{k=1}^K$ such that $z_k \ge T_k^g+{T_{k0}}(p_0,r_0), \forall k \in [K]$, and $z_k \ge [{\bf I}]_{kl} T_{kl}(p_l, r_l), \forall l\in[L], k\in[K]$. Also define $\mathcal{E}^{(r)}_l=\min \left\{\varepsilon_{kl}^{(r)} \big{|} [{\bf I}]_{kl}\neq 0, k\in[K]\right\},~ \forall l\in[L]$ and $\mathcal{E}^{(s)}=\min \left\{\varepsilon_k^{(s)}, k\in[K]\right\}$. Optimization (\ref{OptiProb}) can be simplified as (\ref{OptiProb1})
\begin{equation}\label{OptiProb1}
\begin{aligned}
&\min_{\left\{p_0, r_0, p_{l}, r_{l}, l\in[L], z_{k}, k\in[K]\right\}} & & \sum_{k=1}^K z_k \\
& \text{s.t.} & &  z_k \ge T_k^g+{T_{k0}}(p_0,r_0), \forall k \in [K] \\
&&& z_k \ge [{\bf I}]_{kl} T_{kl}(p_l, r_l), \forall l\in[L], k\in[K] \\
&&& \sum\nolimits_{l=0}^L p_{l}\leq P_{\text{T}},\\
&&& \Phi_r\left(r_l\right) \leq \mathcal{E}^{(r)}_l,~ \forall l\in[L],\\
&&& \Phi_s\left(r_0\right) \leq \mathcal{E}^{(s)}.
\end{aligned}
\end{equation}
It is also straightforward to show that optimization (\ref{OptiProb1}) is convex, as the reconstruction/synthesis distortion/perception curves $\Phi_r\left(r_l\right)$ and $\Phi_s\left(r_0\right)$ are monotonically non-increasing according to the rate-distortion-perception theorem \cite{RDP, RDP2}. This optimization is a constrained non-linear problem, which we solve using sequential quadratic programming (SQP) \cite{NoceWrig06}. Let us define ${\bf z}=[p_0,\cdots,p_L,r_0,\cdots,r_L,z_1,\cdots,z_K]^\top$, and 
\begin{align}
    &f({\bf z})=\sum_{k=1}^K z_k, \\ \nonumber
    &g_k({\bf z})= z_k - T_k^g+{T_{k0}}(p_0,r_0), \forall k \in [K], \\ \nonumber
    &h_{kl}({\bf z})= z_k - [{\bf I}]_{kl} T_{kl}(p_l, r_l), \forall l\in[L], k\in[K], \\ \nonumber
    &i({\bf z})= P_{\text{T}} - \sum\nolimits_{l=0}^L p_{l}, \\ \nonumber
    &j_l({\bf z})= \mathcal{E}^{(r)}_l - \Phi_r\left(r_l\right),~ \forall l\in[L], \\ \nonumber
    &q({\bf z})= \mathcal{E}^{(s)} - \Phi_s\left(r_0\right). 
\end{align}
With SQP, we sequentially calculate Hessian of the Lagrangian function given by $\mathcal{L}({\bf z})= f({\bf z})+ \sum_k \zeta_k g_k({\bf z}) + \sum_k\sum_l \xi_{kl} h_{kl}({\bf z}) + \delta i({\bf z})+ \sum_l \psi_l j_l({\bf z}) + \mu q({\bf z}) $, and linearize the constraints to formulate the approximate quadratic problem (QP) (\ref{OptiProb2}). The optimal update vector ${\bf d}$ for (\ref{OptiProb2}) is obtained by a quasi-newton and line search (LS). Our final algorithm for latency minimization (\ref{OptiProb}) is summarized in \textbf{Algorithm 1}.

\begin{algorithm}[t]
 \captionsetup{font={footnotesize}, singlelinecheck = off,  name={Fig.},labelsep=period}
\caption{\color{black}SQP-based Latency Minimization for Diffusion-based Generative Semantic Multicasting.}
\SetKwInOut{Input}{Input}\SetKwInOut{Output}{Output}

\Input{$[{\bf I}]_{kl}, \forall k,l$, $\{h_k\}_{k=1}^K$, $\{\varepsilon_{kl}^{(r)}\}_{k=1}^K$, $\{\varepsilon_k^{(s)}\}_{k=1}^K$, $\{T_k^g\}_{k=1}^K$, $\{B_l\}_{l=0}^L$, $\Phi_r\left(.\right)$, $\Phi_s\left(.\right)$, $P_T$, $N_0$, $\mathrm{Iter}_\mathrm{max}$, $\mathrm{Tol}$.}

\Output{Optimized $\{p_l^*\}$ and $\{r_l^*\}$.}
\BlankLine

\textbf{Initialize} \textbf{Set:} $i = 0$, \textbf{Randomize:} $\mathbf{z}_0$,$\{\alpha_k\}$, $\{\beta_{kl}\}$, $\{\mu_l\}$, $\delta$, $\sigma$, \textbf{Calculate:} $\mathcal{E}^{(r)}_l=\min \left\{\varepsilon_{kl}^{(r)} \big{|} [{\bf I}]_{kl}\neq 0, k\in[K]\right\},~ \forall l\in[L]$, $\mathcal{E}^{(s)}=\min \left\{\varepsilon_k^{(s)}, k\in[K]\right\}$.\\

\While{$i < \mathrm{Iter}_\mathrm{max}$ \textbf{\bf and} $\|\nabla f(\mathbf{z}_i)\| > \mathrm{Tol}$}{
  - Obtain search direction $\mathbf{d}_i$ from the solution of QP subproblem (\ref{OptiProb2}).

  - LS to determine step size $\eta_i$, $\mathbf{z}_{i+1} = \mathbf{z}_i + \eta_i \mathbf{d}_i.$

  - Stop if $\|\mathbf{d}_i\| < \mathrm{Tol}$.

  - Increment $i = i + 1$.
}

\textbf{End while}

\Return $\mathbf{z}^*=\mathbf{z}_i$, $\{p_l^*\}$, $\{r_l^*\}$.
\end{algorithm}

\begin{align}\label{OptiProb2}
    &\min_{\mathbf{d}} && \nabla f(\mathbf{z})^\top \mathbf{d} + \frac{1}{2} \mathbf{d}^\top \nabla^2_{zz} \mathcal{L}(\mathbf{z}) \mathbf{d}, \\ \nonumber
    &\text{s.t.} && \nabla g_k(\mathbf{z})^\top \mathbf{d} + g_k(\mathbf{z}) \geq 0,~ \forall k \in [K], \\ \nonumber
    & &&\nabla h_{kl}(\mathbf{z})^\top \mathbf{d} + h_{kl}(\mathbf{z}) \geq 0,~ \forall l \in [L],~ k \in [K], \\ \nonumber
    & &&\nabla i(\mathbf{z})^\top \mathbf{d} + i(\mathbf{z}) \geq 0, \\ \nonumber
    & &&\nabla j_l(\mathbf{z})^\top \mathbf{d} + j_l(\mathbf{z}) \geq 0,~ \forall l \in [L], \\ \nonumber
    & &&\nabla q(\mathbf{z})^\top \mathbf{d} + q(\mathbf{z}) \geq 0. 
\end{align}
}

Finally, denoting the minimum sum latency from (\ref{OptiProb}) by $\bar{T}$, the average {\it per-user latency} is given by $T^* = \bar{T}/K$, and the resulting average spectral efficiency is 
\begin{align}
    \lambda^*=\frac{\sum_{k=1}^K R^*_{k0} + \sum_{k=1}^K \sum_{l=1}^L R^*_{kl} [{\bf I}]_{kl}}{B_0+\sum_{l=1}^L B_l [1-\prod_{k=1}^{K}(1-[{\bf I}]_{kl})]},
\end{align}
where the optimal rate $R^*_{k0}$ and $R^*_{kl}$ are calculated according to (\ref{eq:rate1}) and (\ref{eq:rate2}) using the optimally allocated powers from (\ref{OptiProb}), the term $1-\prod_{k=1}^{K}(1-[{\bf I}]_{kl})$ in the denominator enforces that if no user is interested in class $l$, then no signal is transmitted over $B_l$ to save the bandwidth. The average source compression rate for multicasting is $r^*=r_0+\frac{|{\bf \Bar{X}}|}{|{\bf X}|} \sum_{l=1}^{L} r_l$, and the total number of bits transmitted over the wireless channel for multicasting is $|{\bf b}|=\sum_{l=0}^{L} |{\bf b}_l|=r^* \times |{\bf X}|$. Finally, the optimal portion of the power budget allocated to multicasting of the semantic map is given by $\gamma^*=\frac{p_0^*}{\sum_{l=0}^{L} p_l^*}=p_0^*/P_T$, as it is straightforward to investigate that the power constraint is always active in the optimum solution. 

\begin{figure}[t]{}
  \centering
  \includegraphics[width=1.1\linewidth]{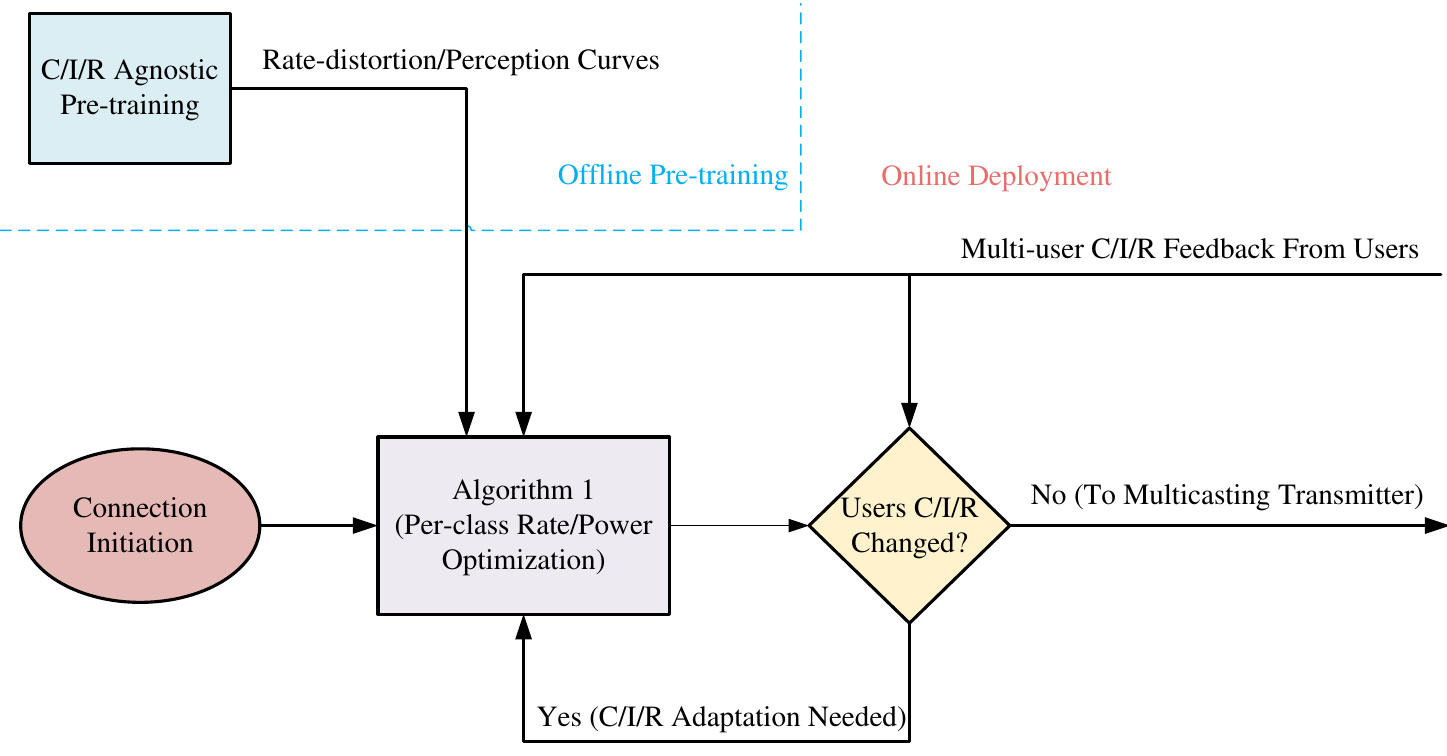}  
   \captionsetup{font={footnotesize}, singlelinecheck = off,  name={Fig.},labelsep=period}
  \caption{\color{black}Separate source-channel coding workflow for generative semantic multicasting.}
  \label{FlowChart}
\end{figure}

{\color{black} \subsection{Separate Source-Channel Coding Workflow} 
Fig. \ref{FlowChart} demonstrates the general flowchart of the proposed generative semantic multicasting framework. The workflow is composed of two phases: 1) Offline Pre-training, and 2) Online Deployment. This two-phase design follows a source-channel coding separation design, where the ``Offline Pre-training" phase of the generative model is agnostic to the user channel conditions $\mathbb{E}|h_k|^2/N_0$, user intents $[\mathbf{I}]_{kl}$, and their distortion/perception requirements $\{\varepsilon_{kl}^{(r)}\}_{k=1}^K$, $\{\varepsilon_k^{(s)}\}_{k=1}^K$. Through pre-training, the rate-distortion/perception curves are derived, which are then used for adaptation to users' channel conditions, intents, and distortion/perception requirements, i.e. C/I/R adaptability, during the ``Online Deployment" phase. The C/I/R adaptation is achieved by running the proposed \textbf{Algorithm 1} during the ``Online Deployment" phase without requiring end-to-end retraining. \textbf{Algorithm 1} inputs the multiuser channel conditions $\mathbb{E}|h_k|^2/N_0$, user intents $[\mathbf{I}]_{kl}$, and their distortion/perception requirements $\{\varepsilon_{kl}^{(r)}\}_{k=1}^K$, $\{\varepsilon_k^{(s)}\}_{k=1}^K$, i.e. C/I/R feedback from users, along with the rate-distortion/perception curves derived during pre-training to adaptively determine the per-class compression rates $\{r_l^*\}$ and transmission powers $\{p_l^*\}$. We should note that the intent feedback overhead is as small as $\left|{\mathbf I}\right|=KL$ bits, which is negligible in practice. Moreover, as \textbf{Algorithm 1} solves a sparsely activated SQP, it has a complexity of $O(I(L+K)^e)$, where $I$ is the number of outer iterations depending on the stopping Tol, and the scaling exponent $e \leq 2$ in the worst case, but here scales much better due to the sparsity structure. In our experiments, we observed an empirical scaling of $1 \leq e \leq 1.5$, which is affordable as this algorithm is run once upon establishment of the multicast communication, and the results stay valid as long as its inputs, i.e. the users C/I/R, remain unchanged.}

\section{Simulation Results} \label{SEC_Result}
\subsection{Reconstruction/Synthesis Rate Distortion/Perception Curves}
{\color{black}In this work, we adopt the MS-SSIM and LPIPS as the distortion and perception metrics to evaluate quality of the reconstructed and synthesized signals at the users, respectively. MS-SSIM (Multi-Scale Structural Similarity Index) is based on the pixel-level differences \cite{SSIM}, while LPIPS (Learned Perceptual Image Patch Similarity) is computed by taking the L2 norm of the differences between feature maps from different layers of a pre-trained VGG network, normalized and weighted by layer importance \cite{LPIPS}. LPIPS is an effective metric for assessing the perceptual quality of images synthesized by generative models, super-resolution, style transfer, etc.} and is known to offer a human-like evaluation.

\begin{figure}[t]
\centering
{\includegraphics[scale=0.4]{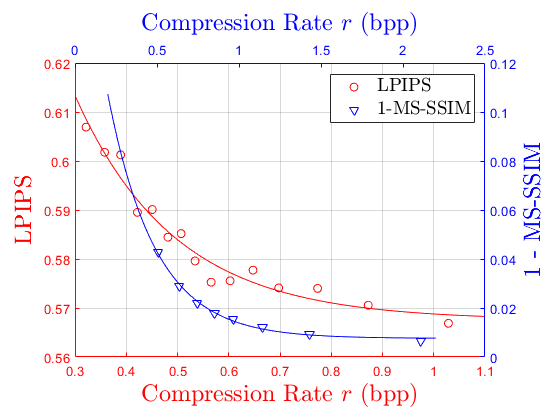}}
 \captionsetup{font={footnotesize}, singlelinecheck = off,   name={Fig.},labelsep=period}
\caption{Reconstruction/Synthesis Distortion/Perception curves.}
\label{fig:LPIPS and MS-SSIM fitting}
\end{figure}

At the users, we pretrain the SDM model explained in subsection \ref{SDMGDM} on the Cityscapes dataset, and carryout extensive simulations to derive the approximate $\Phi_r\left(r\right)=\mathrm{1-MS \scalebox{0.75}[1.0]{\(-\)} SSIM}$ and $\Phi_s\left(r\right)=\mathrm{LPIPS}$ functions via curve fitting. We use the conventional JPEG \cite{JPEG} for adaptive compression of the semantic map and the user-intended signals. Fig. \ref{fig:LPIPS and MS-SSIM fitting} depicts the achieved curve fitting results, which shows that the fitted exponential functions, i.e., $\Phi_r\left(r\right)=0.199 e^{(-3.454 r)} + 0.008$ and $\Phi_s\left(r\right)= 0.214 e^{(-5.14 r)} + 0.566$, very well approximate both metrics with high accuracy. Note that such convex and monotonically non-increasing curves are expected from the rate-distortion-perception theorem \cite{RDP2}.

\subsection{The Simulation Setup}
We consider a generative semantic multicasting setup, where a transmitter equipped with a camera multicasts the street scene images to multiple users based on the multi-user intent. The street scene images are taken randomly from the Cityscapes dataset. Cityscapes is a benchmark for semantic understanding of urban street scenes \cite{Cordts2016Cityscapes}, which includes 8 different semantic categories, namely, construction, object, nature, sky, human, vehicle, flat, and void, and a total of 35 semantic object classes. Each category may include several object classes, e.g., the vehicle category covers ``Bicycle”, ``Motorcycle”, ``Train”, ``Trailer”, ``Caravan”, and ``Car” object classes. Error-free segmentation of the scene images at transmitter is assumed. For now, we assume each user is interested in a random class of the 35, with no two users interested in the same class, i.e. non-overlapped intents. Later in Subsection IV.H, we investigate effects of overlapped intents as well. The wireless transmission parameters are listed in TABLE \ref{tab:wireless}. All results are averaged over 6000 Monte Carlo simulations.

\begin{table}[t]
\centering
 \captionsetup{font={footnotesize}, singlelinecheck = off,   labelsep=period}
	\caption{Parameter settings}
	\begin{center}
		\begin{tabular}{ll}
			\toprule[1.5pt]
			Parameters  & Values \\ \hline
   Distance of user $k$, $d_k$    & $\sim \! \mathcal{U}[150, 550]$ m \\
   Number of users, $K$ & 5, 10, 15, 20 \\
    Dist./Perct. requirement, $(\varepsilon_{kl}^{(r)}, \varepsilon_k^{(s)})$   &(0.02850, 0.58705)\\
        Receivers' generation time, $T_k^g$    & 2 ms \\ 
	Transmit power budget, $P_{\text{T}}$    & 100, 150, 200 mW \\ 
	Path loss at $d_0 =1$ m, $\epsilon_o$      &-30 dB \\ 
	Path loss exponent, $\varphi$   & 3.4 \\ 
	Noise power density, $N_0$                 & -174 dBm/Hz \\ 
	Bandwidth, $B_0=B_l$                         & 1~MHz \\
			\toprule[1.5pt]
		\end{tabular}
	\end{center}
	\label{tab:wireless}
  
\end{table}

\begin{table}[t]
\centering
 \captionsetup{font={footnotesize}, singlelinecheck = off,   labelsep=period}
\caption{The average source compression rate $r^*$, and the total number of transmitted bits $|{\bf b}|$ for various number of users.}  
\label{tab:bits}  
\begin{tabular}{|c|c|c|c|c|c|c|}
\hline
$K$ & 1  & 2 & 3 & 5 & 10 & 15 \\ \hline\hline
$r^*$ & 0.51698 & 0.58279 & 0.64859 & 0.7802 & 1.10922 & 1.43823 \\ \hline
$|{\bf b}|$ & 67762 & 76387 & 85012 & 102262 & 145387 & 188512 \\ \hline
\end{tabular}
\end{table}

\subsection{Comparison with Non-Generative and Full Generative Intent-Unaware Multicasting Benchmarks}
{\color{black}We compare our proposed intent-aware generative semantic multicasting framework with the following two benchmarks: 
     
     \textit{1) Non-Generative intent-unaware Multicasting (NGM):} In this benchmark, the transmitter compresses the source signal at rate $r_{NGM}=\Phi_r^{-1}\left( \min \left\{\varepsilon_{kl}^{(r)} \big{|} k\in[K], l\in[L]\right\} \right),$ and multicasts it to all the users over shared resources; 
     
     \textit{2) Full Generative intent-unaware Multicasting (FGM) \cite{SDM2025}:} In this benchmark, the transmitter compresses the semantic map at rate $r_{FGM}=\Phi_s^{-1}\left(\min \left\{\varepsilon_k^{(s)}, k\in[K]\right\}\right)$ and multicasts it to all users over shared resources, where each user then carries out SDM according to \cite{SDM2025}, to locally synthesize a signal utilizing its received semantic map.}

\begin{figure}[t]
\centering
\begin{subfigure}[h]{0.49\columnwidth}
  \centering
  \includegraphics[width=\linewidth]{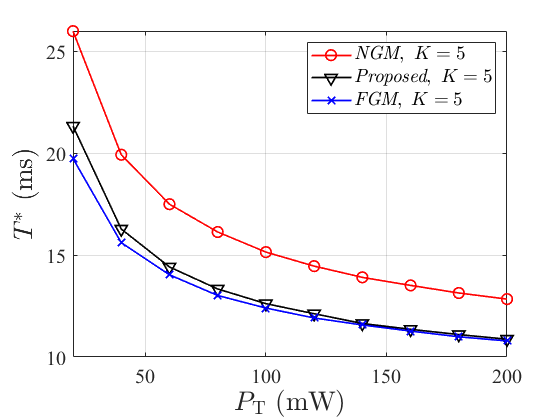}  
  \caption{}
  \label{fig.9(a)}
\end{subfigure}
\begin{subfigure}[h]{0.49\columnwidth}
  \centering
  \includegraphics[width=\linewidth]{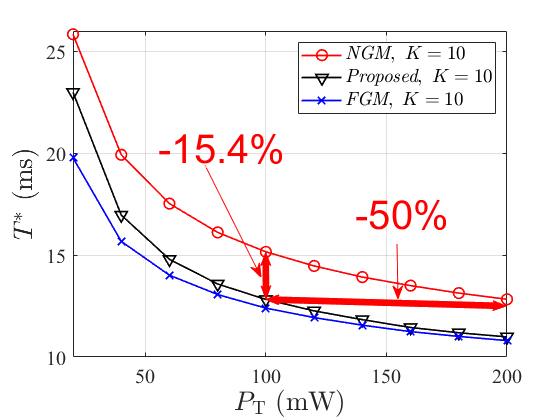}  
  \caption{}
  \label{fig.9(b)}
\end{subfigure}
\begin{subfigure}[h]{0.49\columnwidth}
  \centering
  \includegraphics[width=\linewidth]{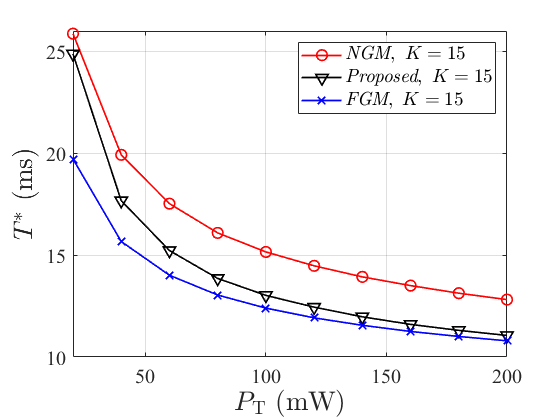}  
  \caption{}
  \label{fig.9(c)}
\end{subfigure}
\begin{subfigure}[h]{0.49\columnwidth}
  \centering
  \includegraphics[width=\linewidth]{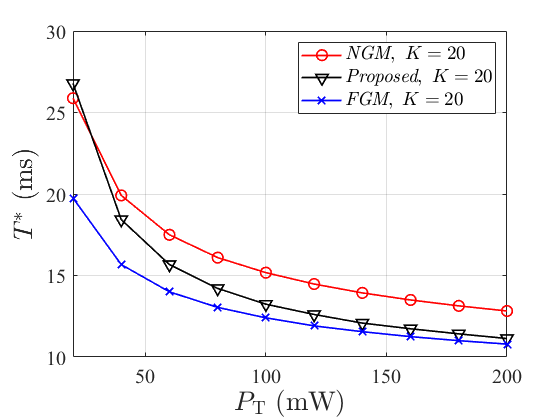}  
  \caption{}
  \label{fig.9(d)}
\end{subfigure}
 \captionsetup{font={footnotesize}, singlelinecheck = off,  name={Fig.},labelsep=period}
\caption{Per-user latency comparison between the proposed generative semantic multicasting framework and the benchmarks, for (a) $K=5$, (b) $K=10$, (c) $K=15$ (d) $K=20$.}
\label{fig:9}
 
\end{figure}

In  TABLE \ref{tab:bits}, we report the average multicasting semantic compression rate $r^*$ (bpp), and the resulting total number of bits transmitted over the wireless channel $|{\bf b}|=\sum_{l=0}^{L} |{\bf b}_l|$. As expected, the $r^*$ and $|{\bf b}|$ values increase, with increasing number of users $K$. On the other hand, from the rate-distortion/perception curves in Fig. \ref{fig:LPIPS and MS-SSIM fitting}, it is straightforward to show that the compression rate for the FGM and NGM benchmarks is $r_{NGM}=\Phi_r^{-1}(0.02850)=0.65804$ and $r_{FGM}=\Phi_s^{-1}(0.58705)=0.45118$, respectively. Comparing these values with the compression rate achieved by our proposed framework as reported in TABLE \ref{tab:bits}, we see that $r^*<r_{NGM}$ for $K\leq3$, while for $K>3$ we have $r^*>r_{NGM}$. This shows that our proposed framework achieves a smaller compression rate for smaller $K\leq3$ values, thereby transmitting fewer bits over the wireless channel. In comparison with FGM, we have $r^*>r_{FGM}$, i.e. our proposed framework increases the total number of bits communicated over the wireless channel as expected, due to additional transmission of the user-intended signals. 

In Fig. \ref{fig:9} we compare the per-user latency curves of the proposed scheme with the benchmarks for various number of users $K$, where $(\varepsilon_{kl}^{(r)}, \varepsilon_k^{(s)})=(0.02850, 0.58705)$. Our proposed framework significantly reduces the per-user latency in comparison with NGM. Referring to TABLE \ref{tab:bits}, despite increasing the total number of bits for $K>3$, our proposed approach reduces the per-user latency in comparison with the conventional NGM even for large $K$, thereby improving scalability. It becomes clear that the amount of multicasted information, i.e. number of multicasted bits, dominates the per-user latency. This is expected due to the fact that among the $K$ users, the user with the worst channel becomes the bottleneck in multicasting, thereby increasing the latency. Despite increasing the number of transmitted bits, our proposed framework reduces the number of multicasted bits, consequently reducing the per-user latency for a fixed total spectrum resources and transmit power budget. {\color{black}According to Fig. \ref{fig:9}, our proposed generative multicasting framework reduces the per-user latency by 16.7\%, 15.4\%, 14.1\%, and 12.8\% in comparison with NGM for $K=5, 10, 15, 20$, respectively, when the transmission power budget is fixed at $P_{\text{T}}=100$mW. Equivalently, our proposed framework reduces the transmission power required to achieve a target per-user latency of $13$ms, by 55.6\%, 50.0\%, 44.4\%, and 44.4\%, for $K=5, 10, 15, 20$, respectively.} Finally, in comparison with FGM, our proposed framework only slightly increases the latency, which is due to transmission of the user-intended signals over orthogonal resources.

In Fig. \ref{fig.Spectral efficiency} we compare the achieved spectral efficiency curves of the proposed scheme with the benchmarks for various number of users $K$. It should be noted that FGM and NGM benchmarks achieve similar spectral efficiency as they both multicast the signal to all users over the same spectrum resources and power budget. As depicted in Fig.  \ref{fig.Spectral efficiency}, for larger number of users i.e. $K=15, 20$, and larger transmit power budget i.e.  $P_{\text{T}} \geq 100$mW, our proposed framework achieves a higher spectral efficiency.

\begin{figure}[t]{}
  \centering
  \includegraphics[width=.75\linewidth]{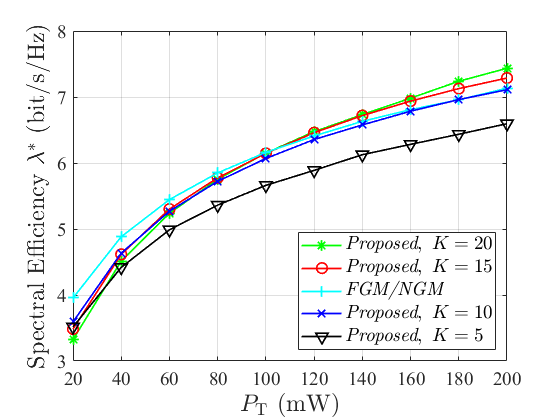}  
   \captionsetup{font={footnotesize}, singlelinecheck = off, name={Fig.},labelsep=period}
  \caption{Spectral efficiency comparison between the proposed generative semantic multicasting framework and the benchmarks.}
  \label{fig.Spectral efficiency}
\end{figure}

\begin{figure*}[t]
\centering
\begin{subfigure}[h]{0.3\linewidth}
  \centering
  \includegraphics[width=\linewidth]{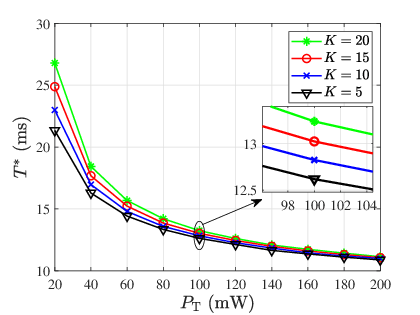}  
  \caption{}
  \label{fig.3(a)}
\end{subfigure}
\begin{subfigure}[h]{0.3\linewidth}
  \centering
  \includegraphics[width=\linewidth]{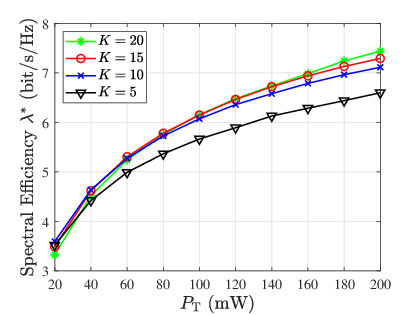}  
  \caption{}
  \label{fig.3(b)}
\end{subfigure}
\begin{subfigure}[h]{0.3\linewidth}
  \centering
  \includegraphics[width=\linewidth]{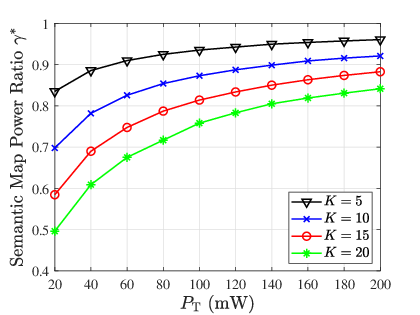}  
  \caption{}
  \label{fig.3(c)}
\end{subfigure}
 \captionsetup{font={footnotesize}, singlelinecheck = off,  name={Fig.},labelsep=period}
\caption{The achieved average (a) per-user latency, (b) spectral efficiency, and (c) power ratio of the semantic map versus the transmit power budget, for various numbers of users.}
\label{fig:3}
\end{figure*}

\begin{figure*}[h!]
\centering
\begin{subfigure}[h]{0.3\linewidth}
  \centering
  \includegraphics[width=\linewidth]{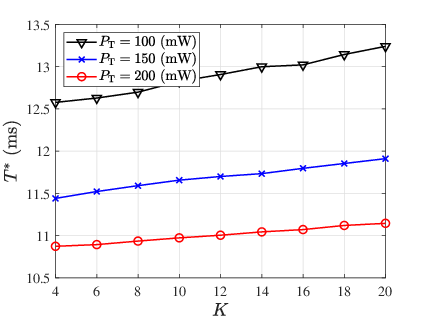}  
  \caption{}
  \label{fig.4(a)}
\end{subfigure}
\begin{subfigure}[h]{0.3\linewidth}
  \centering
  \includegraphics[width=\linewidth]{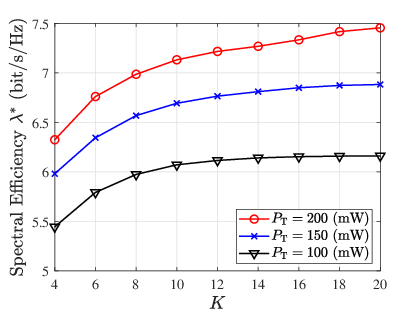}  
  \caption{}
  \label{fig.4(b)}
\end{subfigure}
\begin{subfigure}[h]{0.3\linewidth}
  \centering
  \includegraphics[width=\linewidth]{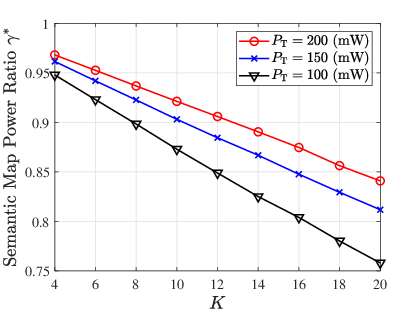}  
  \caption{}
  \label{fig.4(c)}
\end{subfigure}
 \captionsetup{font={footnotesize}, singlelinecheck = off,  name={Fig.},labelsep=period}
\caption{The achieved average (a) per-user latency, (b) spectral
efficiency, and (c) power ratio of the semantic map
versus the number of users, for different maximum power budgets.}
\label{fig:4}
\end{figure*}

\subsection{Effects of the Transmit Power Budget}
In Fig. \ref{fig:3} we study effects of the transmit power budget $P_{\text{T}}$ for various number of users $K=5, 10, 15, 20$ on the achieved average per-user latency $T^*$, the spectral efficiency $\lambda^*$, and the optimum ratio of the total power allocated to multicasting of the semantic map $\gamma^*$ from optimization (\ref{OptiProb}). The average per-user latency decreases and the spectral efficiency increases with increasing $P_{\text{T}}$. Moreover, as the transmit power budget increases, a larger portion of the total power is allocated to multicasting of the semantic map.

\subsection{Effects of the Number of Users}
In Fig. \ref{fig:4} we study effects of increasing the number of users $K$ to investigate scalability of the proposed generative multicasting framework for different transmit power budget $P_{\text{T}}=100, 150, 200$ mW. Both the average per-user latency and the spectral efficiency increase with increasing number of users $K$. Moreover, as the number of users increases, a smaller portion of the total power budget is allocated to multicasting of the semantic map.

\subsection{\color{black}Effects of the Generation Latency}
{\color{black}Despite their success in synthesizing high quality natural signals, diffusion models have a higher generation latency compared to other generative models. There is a great deal of ongoing research to reduce the generation time of these models and facilitate their deployment on mobile devices. More and more computationally efficient diffusion models are introduced every day. The generation latency depends on the number of FLOPs required to run the adopted diffusion model denoted by $N^F$, and the computation power of the processor at each user, i.e. $P^F_k$ in FLOPs per Second. The resulting computation latency for user $k$ is thereby given by $T_k^g=N^F/P^F_k$.

\begin{table}[t]
\centering
 \captionsetup{font={footnotesize}, singlelinecheck = off, labelsep=period}
\caption{\color{black}The generation time for diffusion models on typical desktop/device AI processors.}  
\label{tab:FLOPS}  
\begin{tabular}{|cc|cc|}
\hline
\multicolumn{2}{|c|}{\begin{tabular}[c]{@{}c@{}}Diffusion \cite{LightDiff}\\ 2.87 TFLOPs\end{tabular}} & \multicolumn{2}{c|}{\begin{tabular}[c]{@{}c@{}}Mobile Diffusion-Lite \cite{MobDiff}\\ 153 GFLOPs\end{tabular}} \\ \hline\hline
\multicolumn{1}{|c|}{\begin{tabular}[c]{@{}c@{}}NVIDIA A100\\ 312 TFLOPS\end{tabular}} & \begin{tabular}[c]{@{}c@{}}NVIDIA H100\\ 989 TFLOPS\end{tabular} & \multicolumn{1}{c|}{\begin{tabular}[c]{@{}c@{}}Apple A18\\ 35 TFLOPS\end{tabular}} & \begin{tabular}[c]{@{}c@{}}Apple M4\\ 38 TFLOPS\end{tabular} \\ \hline
\multicolumn{1}{|c|}{9.2 ms} & 2.9 ms & \multicolumn{1}{c|}{4.4 ms} & 4.0 ms \\ \hline
\end{tabular}
\end{table}

\begin{figure}[t]
\centering
\begin{subfigure}[h]{0.49\columnwidth}
  \centering
  \includegraphics[width=\linewidth]{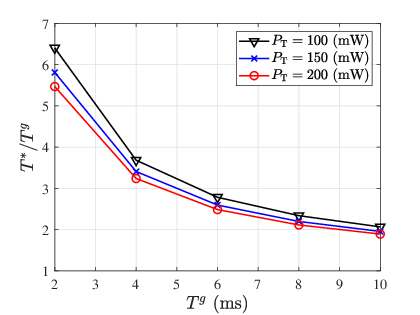}  
  \caption{}
  \label{fig:ComputationTime(a)}
\end{subfigure}
\begin{subfigure}[h]{0.49\columnwidth}
  \centering
  \includegraphics[width=\linewidth]{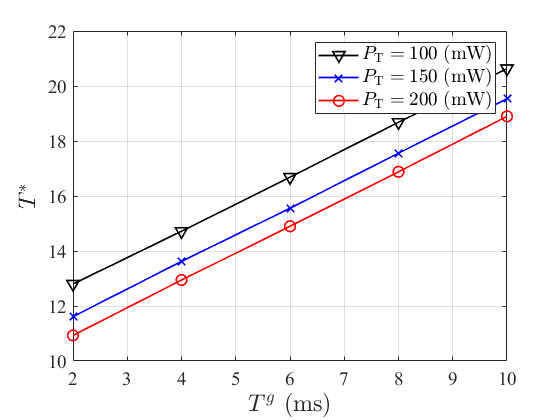}  
  \caption{}
  \label{fig:ComputationTime(b)}
\end{subfigure}
 \captionsetup{font={footnotesize}, singlelinecheck = off,name={Fig.},labelsep=period}
\caption{\color{black}Effects of the generation time and the corresponding communication-computation tradeoffs, $K=10$.}
\vspace{-5mm}
\label{fig:ComputationTime}
\end{figure}

In our proposed, generative semantic multicasting scheme, the receivers may either be conventional users equipped with high performance AI processors (e.g. NVIDIA A100 or H100), or mobile users equipped with mobile AI processors (e.g. Apple A18 or M4). On the high performance processors, where we can afford more FLOPs, we use the lightweight diffusion model \cite{LightDiff} to maintain the generation quality, while on the mobile processors, we use the recently released MobileDiffusion \cite{MobDiff} to enable running the model on-device. In TABLE \ref{tab:FLOPS}, we report the resulting generation latency $T_k^g$ for each of these cases. It is concluded that, depending on the deployed model and the device processor, the generation latency is in the range $T_k^g \in [2,10]$ ms. Hence, we set $T_1^g=\cdots=T_K^g=T^g$ and plot the ratio of the average per-user latency over the generation latency, i.e. $T^*/T^g$, for various $T^g$ values in this range when $K=10$, in Fig. \ref{fig:ComputationTime}a. This ratio decreases when the generation latency or transmit power budget increases, demonstrating that the generation latency dominates the total latency with increasing $T_k^g$ and $P_T$. Fig. \ref{fig:ComputationTime}b studies the corresponding communication-computation tradeoffs. For example, according to Fig. \ref{fig:ComputationTime}b, a 2 ms increase in the generation latency (6ms $->$ 8ms), can be mitigated by increasing the transmit power budget $P_T$ (100mw $->$ 200mw) to keep the total latency unchanged at $\approx 17$ms.}

\begin{figure}[t]
\centering
\begin{subfigure}[h]{0.49\columnwidth}
  \centering
  \includegraphics[width=\linewidth]{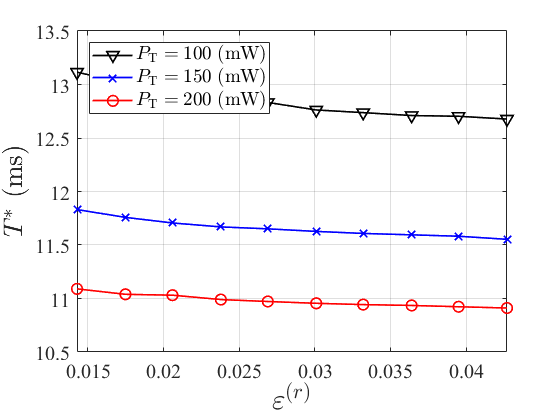}  
  \caption{}
  \label{fig.6(b)}
\end{subfigure}
\begin{subfigure}[h]{0.49\columnwidth}
  \centering
  \includegraphics[width=\linewidth]{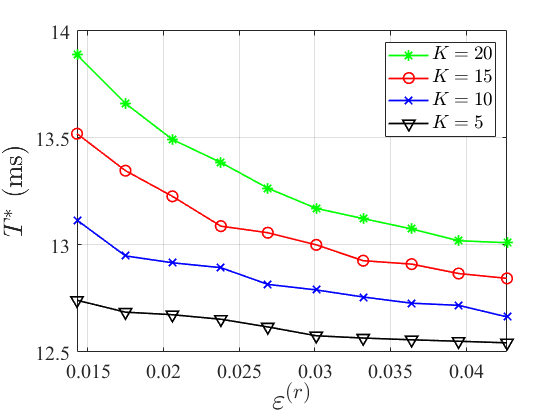}  
  \caption{}
  \label{fig.6(d)}
\end{subfigure}
\begin{subfigure}[h]{0.49\columnwidth}
  \centering
  \includegraphics[width=\linewidth]{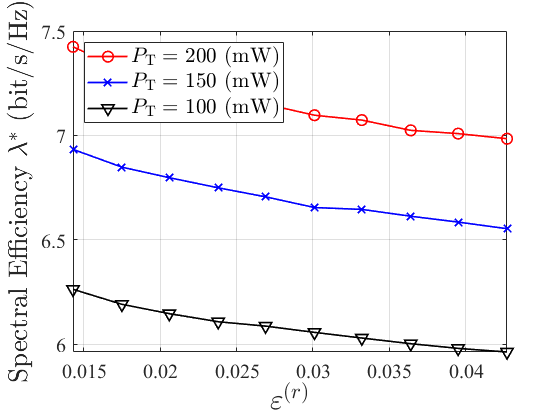}  
  \caption{}
  \label{fig.6(e)}
\end{subfigure}
\begin{subfigure}[h]{0.49\columnwidth}
  \centering
  \includegraphics[width=\linewidth]{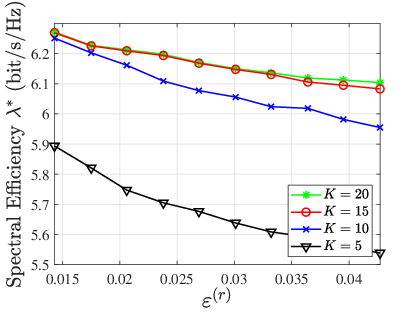}  
  \caption{}
  \label{fig.6(f)}
\end{subfigure}
\begin{subfigure}[h]{0.49\columnwidth}
  \centering
  \includegraphics[width=\linewidth]{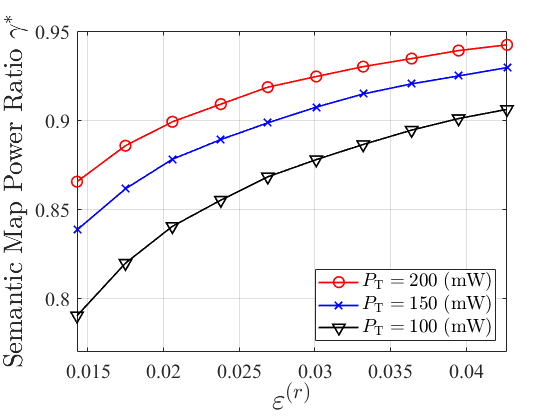}  
  \caption{}
  \label{fig.6(a)}
\end{subfigure}
\begin{subfigure}[h]{0.49\columnwidth}
  \centering
  \includegraphics[width=\linewidth]{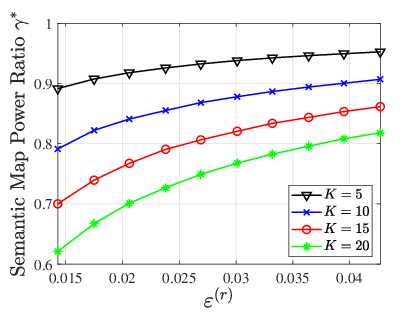}  
  \caption{}
  \label{fig.6(c)}
\end{subfigure}
 \captionsetup{font={footnotesize}, singlelinecheck = off,  name={Fig.},labelsep=period}
\caption{Effects of the reconstruction quality
requirement $\varepsilon^{(r)}_{kl}$.}
\label{fig:6}
\end{figure}

\subsection{Effects of the Reconstruction/Synthesis Quality Requirements}
In Fig. \ref{fig:6}, we study the effects of the reconstruction quality requirement $\varepsilon_{kl}^{(r)}$, on the achieved average per-user latency $T^*$, the spectral efficiency $\lambda^*$, and the optimum power ratio for multicasting of the semantic map $\gamma^*$. As the reconstruction quality requirement is relaxed, the per-user latency and the spectral efficiency is reduced, while a larger portion of the total power budget is allocated to multicasting of the semantic map.

\begin{figure}[t]
\centering
\begin{subfigure}[h]{0.49\columnwidth}
  \centering
  \includegraphics[width=\linewidth]{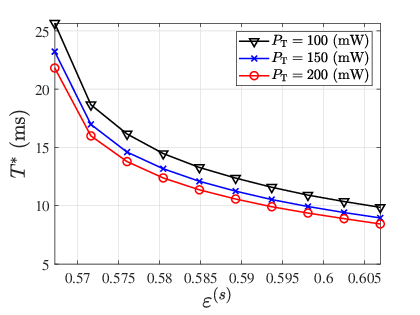}  
  \caption{}
  \label{fig.5(a)}
\end{subfigure}
\begin{subfigure}[h]{0.49\columnwidth}
  \centering
  \includegraphics[width=\linewidth]{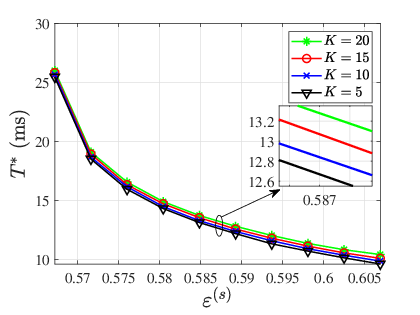}  
  \caption{}
  \label{fig.5(b)}
\end{subfigure}
\begin{subfigure}[h]{0.49\columnwidth}
  \centering
  \includegraphics[width=\linewidth]{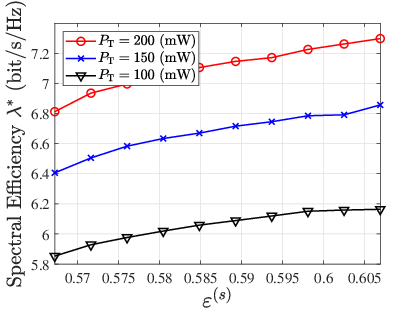}  
  \caption{}
  \label{fig.5(c)}
\end{subfigure}
\begin{subfigure}[h]{0.49\columnwidth}
  \centering
  \includegraphics[width=\linewidth]{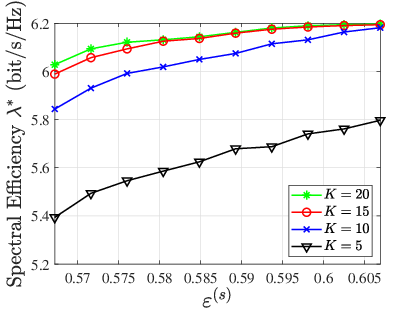}  
  \caption{}
  \label{fig.5(d)}
\end{subfigure}
\begin{subfigure}[h]{0.49\columnwidth}
  \centering
  \includegraphics[width=\linewidth]{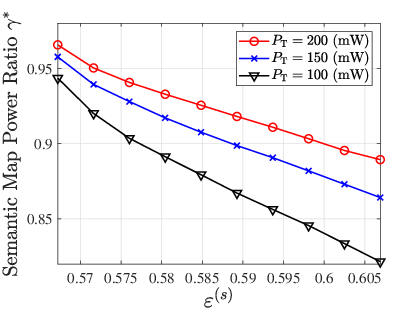}  
  \caption{}
  \label{fig.5(e)}
\end{subfigure}
\begin{subfigure}[h]{0.49\columnwidth}
  \centering
  \includegraphics[width=\linewidth]{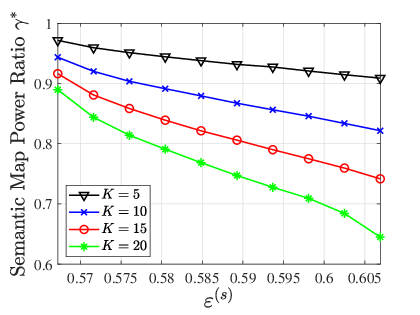}  
  \caption{}
  \label{fig.5(f)}
\end{subfigure}
 \captionsetup{font={footnotesize}, singlelinecheck = off,  name={Fig.},labelsep=period}
\caption{Effects of the synthesis quality
requirement $\varepsilon^{(s)}_{k}$.}
\label{fig:5}
\end{figure}

In Fig. \ref{fig:5}, we study effects of the synthesis quality requirement $\varepsilon_{k}^{(s)}$, on the achieved average per-user latency $T^*$, the spectral efficiency $\lambda^*$, and the optimum power ratio for multicasting of the semantic map $\gamma^*$. As the synthesis requirement is relaxed, the per-user latency is reduced, and the spectral efficiency is increased, while a smaller portion of the total power budget is allocated to multicasting of the semantic map.

\subsection{Effects of Overlapped Intents}
So far, we assumed that no users are interested in the same class, i.e. non-overlapped intents. In this subsection, we demonstrate the further gains achievable when some users may be interested in the same classes, i.e. overlapped intents. To this end, we compare the following three benchmarks assuming $K=10$ users: a) Each of the 10 users is interested in 2 non-overlapped classes making a total of 20 classes transmitted over orthogonal channels apart from the semantic map; b) Each of the 10 users are interested in 2 classes, but the classes are overlapped making up a total of 10 classes transmitted over orthogonal channels apart from the semantic map; c) Each of the 10 users are interested in a single class without overlap (similar to the previous subsections), making a total of 10 classes transmitted over orthogonal channels apart from the semantic map. The intent graphs for these three benchmarks are demonstrated in Fig. \ref{fig:Intent}, where each connection between a user and a class shows that user is interested in that semantic class.

It is straightforward to see that the achieved compression rate $r^*$ is $1.76726$, $1.10922$, and $1.10922$, and the total bandwidth used for the three benchmarks is 21, 11, and 11 MHz for the three benchmarks (a), (b), and (c), respectively. We plot the per-user latency and the corresponding spectral efficiency for the three benchmarks in Fig. \ref{fig:10}. In comparison between (a) and (b), while each user is interested in the same number of classes, i.e. 2, in both benchmarks, the spectral efficiency is increased and the latency, the total number of bits transmitted and the bandwidth used, is decreased in (b) due to overlapped intents. When using the same bandwidth to transmit the same number of bits, i.e. in comparison between benchmarks (b) and (c), the overlapped intents, i.e. (b), improves the spectral efficiency while only slightly increasing the latency. These results show the further benefits achievable with overlapped intents.

\begin{figure}[t]
\centering
\includegraphics[width=.85\linewidth]{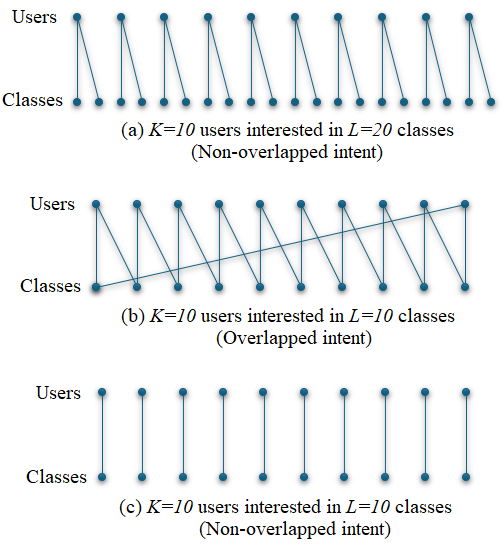}
 \captionsetup{font={footnotesize}, singlelinecheck = off,  name={Fig.},labelsep=period}
\caption{The intent graphs for the considered benchmarks.}
\label{fig:Intent}
\end{figure}

\begin{figure}[t]
\centering
\begin{subfigure}[h]{0.49\columnwidth}
  \centering
  \includegraphics[width=\linewidth]{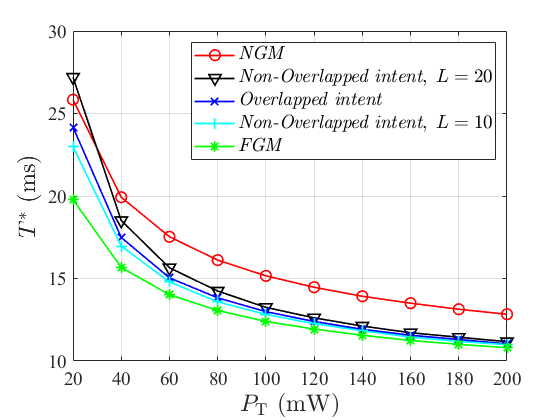}  
  \caption{}
  \label{fig.10(a)}
\end{subfigure}
\begin{subfigure}[h]{0.49\columnwidth}
  \centering
  \includegraphics[width=\linewidth]{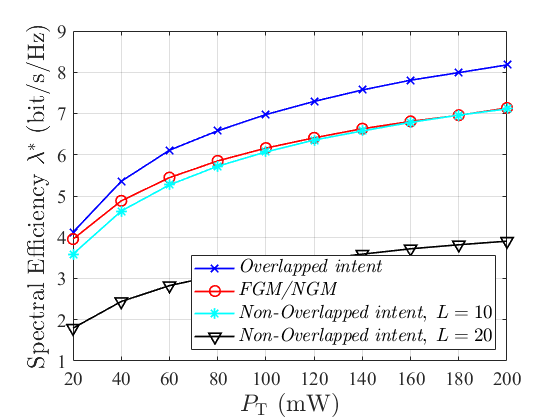}  
  \caption{}
  \label{fig.10(b)}
\end{subfigure}
 \captionsetup{font={footnotesize}, singlelinecheck = off,  name={Fig.},labelsep=period}
\caption{Effects of overlapped intents on the achieved average (a) per-user latency, (b) spectral
efficiency, $P_T=100$ mW and $K=10$.}
\label{fig:10}
 
\end{figure}

\subsection{Visual Quality of the Retrieved Signals}
    In this subsection, we demonstrate the visual quality of the retrieved signals for various reconstruction/synthesis quality requirements and their effect on the average per-user latency achieved by our proposed framework. {\color{black} To demonstrate the effectiveness and generalizability of our proposed framework, we provide simulations both on the Cityscapes and COCO-Stuff datasets.}
\subsubsection{Visual Quality on Cityscapes Street Scenes Dataset}
To this end, we illustrate the visual quality for two typical street scenes in Fig. \ref{fig:VQ1} and \ref{fig:VQ2}, where the transmit power budget is $P_{\text{T}}=100$ mw, the number of users is $K=10$, and the reconstruction quality requirements is fixed at $(\varepsilon_{kl}^{(r)}, \varepsilon_k^{(s)})=(0.02850, 0.58705)$ for all users and classes. The figures show the original scene image, a sample image synthesized locally by the semantic diffusion model, and the final retrieved signals that is partially reconstructed and partially synthesized for three users. It should be noted that, as the random seed and consequently the randomized instance of noise input to the diffusion model is different at various users, the image synthesized locally by semantic guidance is different at various users. In both scenes, user 1 is interested in the ``Building" class, while users 2 and 3 are interested in the ``Car" and ``Road" class, respectively. In this case, the visual quality is rather high and the average per-user latency achieved is $T^*=12.824$ ms. Next, we slightly relax the reconstruction/synthesis quality requirements fixing it at $(\varepsilon_{kl}^{(r)}, \varepsilon_k^{(s)})=(0.06781, 0.63050)$ for all users and classes, and provide the visual quality for the same two street scenes in Fig. \ref{fig:VQ3} and \ref{fig:VQ4}. The visual quality, while still acceptable, does show some degradation in both scenes, i.e., the reconstructed signal portions are slightly pixelated, and the synthesized portions have slightly lost color contrast. Moreover, the average per-user latency is now reduced to $T^*=7.5635$ ms. These results demonstrate high visual quality of the proposed generative semantic multicasting framework, while allowing rather simple adaptation to various user requirements under the prevailing channel conditions.

\begin{figure}[t]
\centering
\includegraphics[scale=0.43]{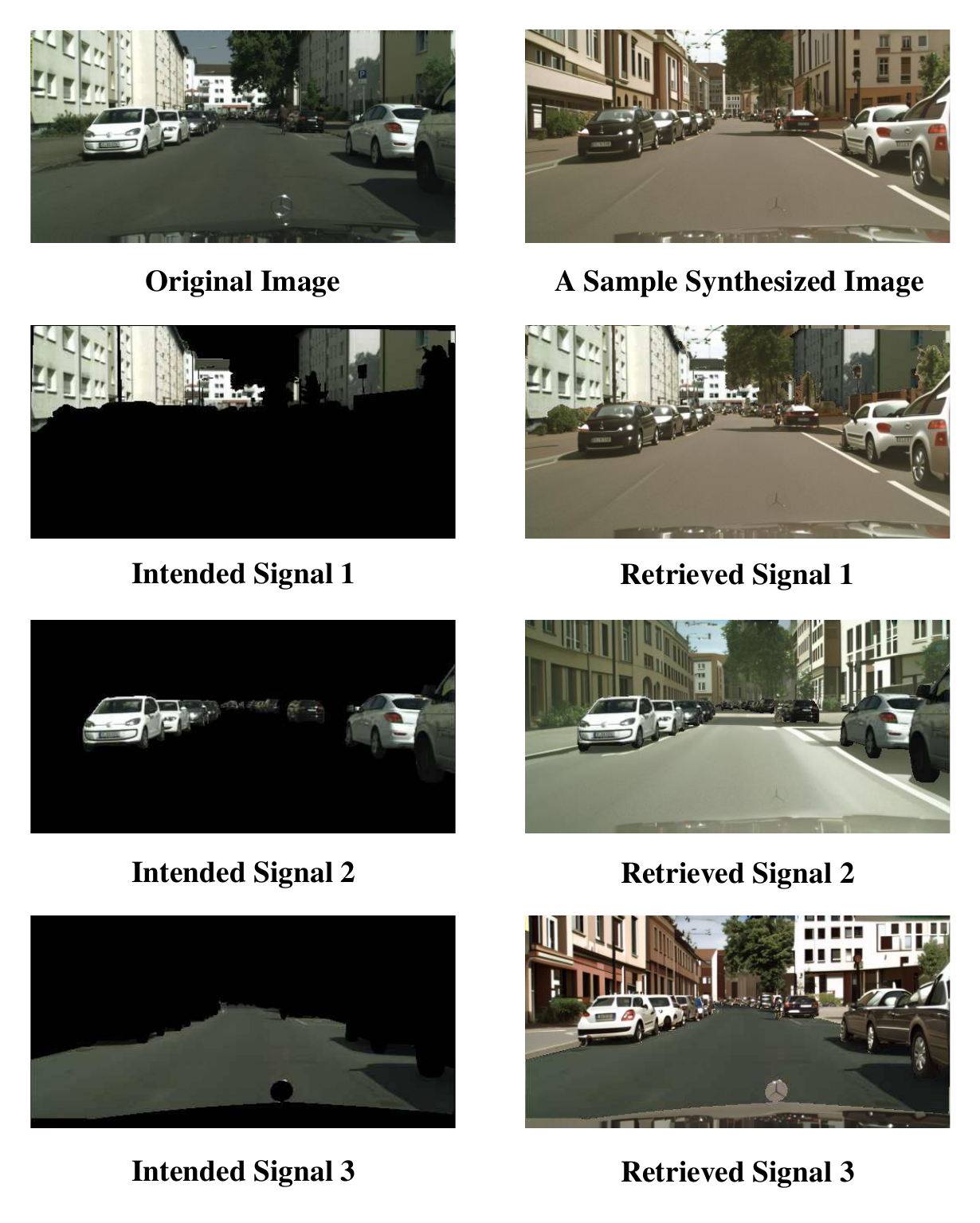}
 \captionsetup{font={footnotesize}, singlelinecheck = off,  name={Fig.},labelsep=period}
\caption{Visual quality of the proposed generative semantic multicasting framework for typical street scene 1, where $(\varepsilon_{kl}^{(r)}, \varepsilon_k^{(s)})=(0.02850, 0.58705)$, $T^*=12.824$ $\rm{ms}$, $\lambda^* = 6.0689$.}
\label{fig:VQ1}
\end{figure}

\begin{figure}[t]
\centering
\includegraphics[scale=0.43]{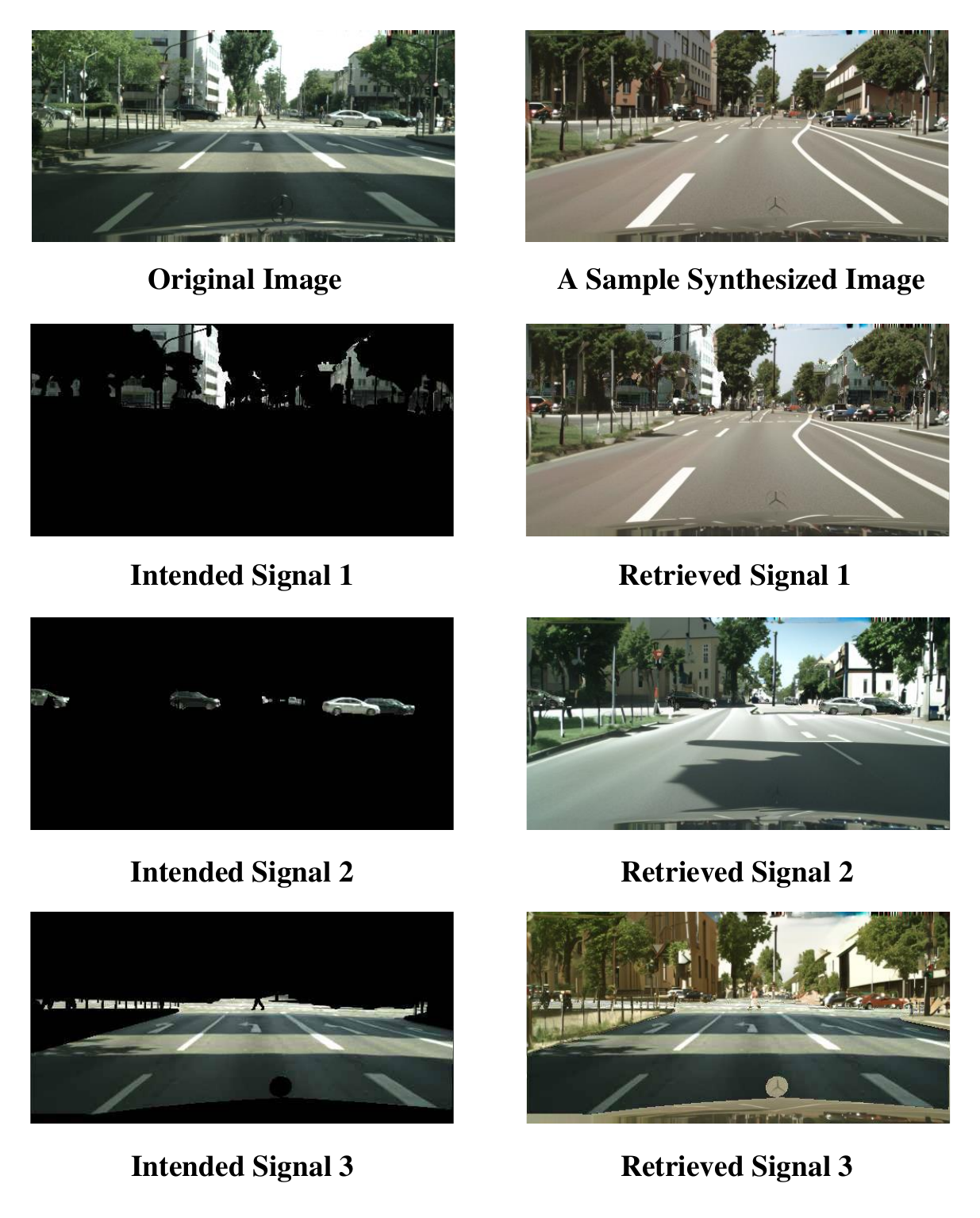}
 \captionsetup{font={footnotesize}, singlelinecheck = off,  name={Fig.},labelsep=period}
\caption{Visual quality of the proposed generative semantic multicasting framework for typical street scene 2, where $(\varepsilon_{kl}^{(r)}, \varepsilon_k^{(s)})=(0.02850, 0.58705)$, $T^*=12.824$ $\rm{ms}$, $\lambda^* = 6.0689$.}
\label{fig:VQ2}
\end{figure}

{\color{black} Next, we provide visual quality comparisons with GAN-based synthesis of the non-intended classes in Fig. 18. We replace the diffusion model with a state-of-the-art {pretrained GAN model for semantic guided synthesis \cite{oasisgan},} while all other simulation parameters remain the same. As observed, synthesis with the diffusion model provides photrealistic sharp outputs with better diversity of colors and realism. GAN-based synthesis is blurrier and exhibits noticeable distortion on the class edges, which significantly degrades the overall visual quality.}

\subsubsection{Visual Quality on COCO-Stuff Dataset}
{\color{black} To demonstrate effectiveness and generalizability of our proposed framework, we conducted simulations on the additional COCO-Stuff dataset, while keeping the simulation parameters and diffusion model unchanged. COCO-Stuff augments the original COCO dataset with dense annotations for 171 object and stuff classes, supporting diverse semantic segmentation tasks. The visualquality results are provided in Fig. \ref{fig:VQ_COCO}, where for both (a) and (b) scenes, the reconstruction/synthesis quality requirements is fixed at $(\varepsilon_{kl}^{(r)}, \varepsilon_k^{(s)})=(0.02850, 0.58705)$ for all users and classes. Each figure shows the original scene image, the corresponding semantic map, and the final retrieved signals for 2 of the users with different intents, which are partially reconstructed and partially synthesized. In scene (a), user 1 is interested in the ``Train" class, while user 2 is interested in the `Road" class. In scene (b), user 1 is interested in the ``Bus" class, while user 2 is interested in the ``Road" class. These results demonstrate high visual quality of the proposed generative semantic multicasting framework, and its effectiveness and generalizability to different datasets and application scenarios.}

\begin{figure}[t]
\centering
\includegraphics[scale=0.43]{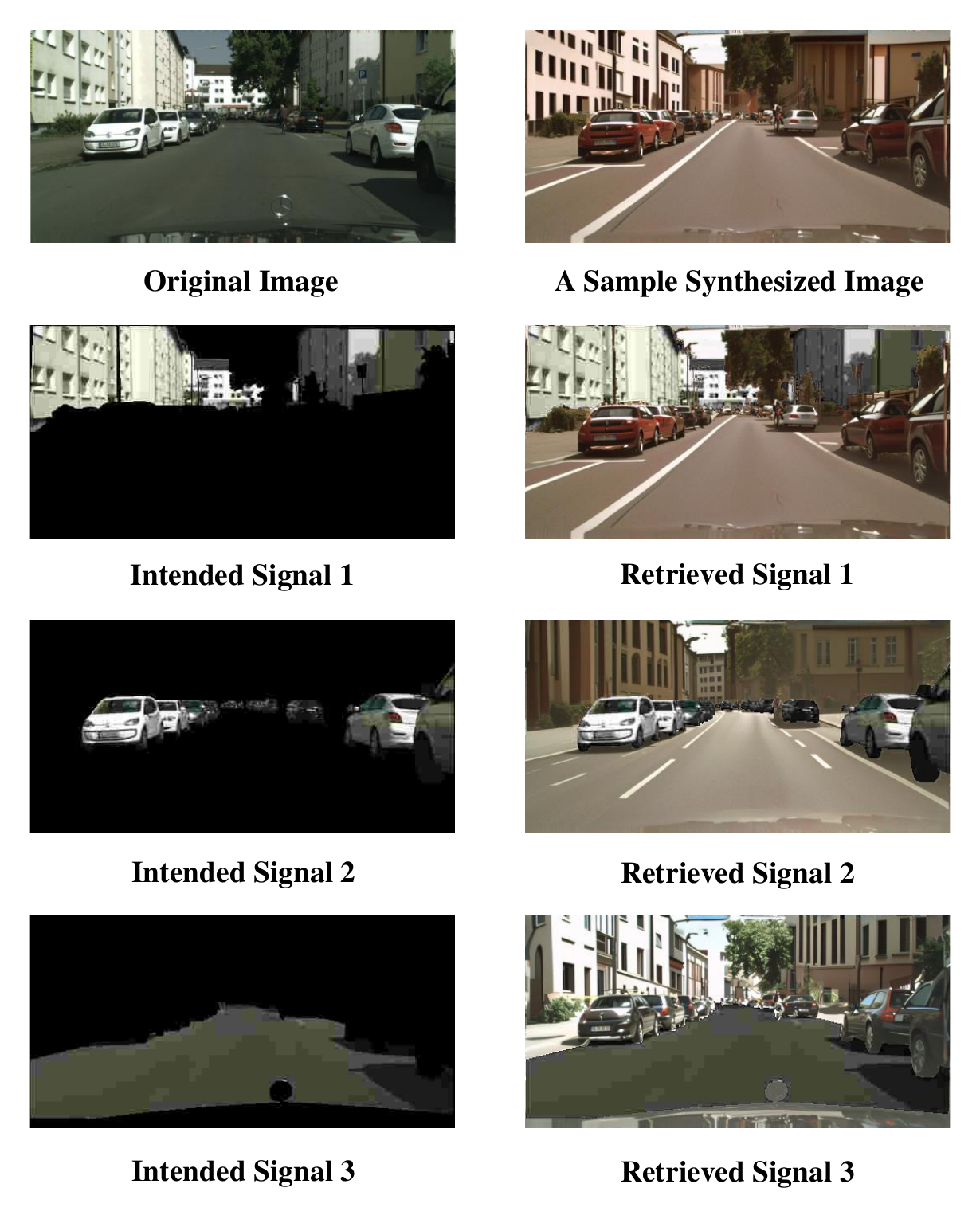}
 \captionsetup{font={footnotesize}, singlelinecheck = off,  name={Fig.},labelsep=period}
\caption{Visual quality of the proposed generative semantic multicasting framework for typical street scene 1, where $(\varepsilon_{kl}^{(r)}, \varepsilon_k^{(s)})=(0.06781, 0.63050)$, $T^*=7.5635$ $\rm{ms}$, $\lambda^* = 6.0154$.}
\label{fig:VQ3}
\end{figure}

\begin{figure}[t]
\centering
\includegraphics[scale=0.43]{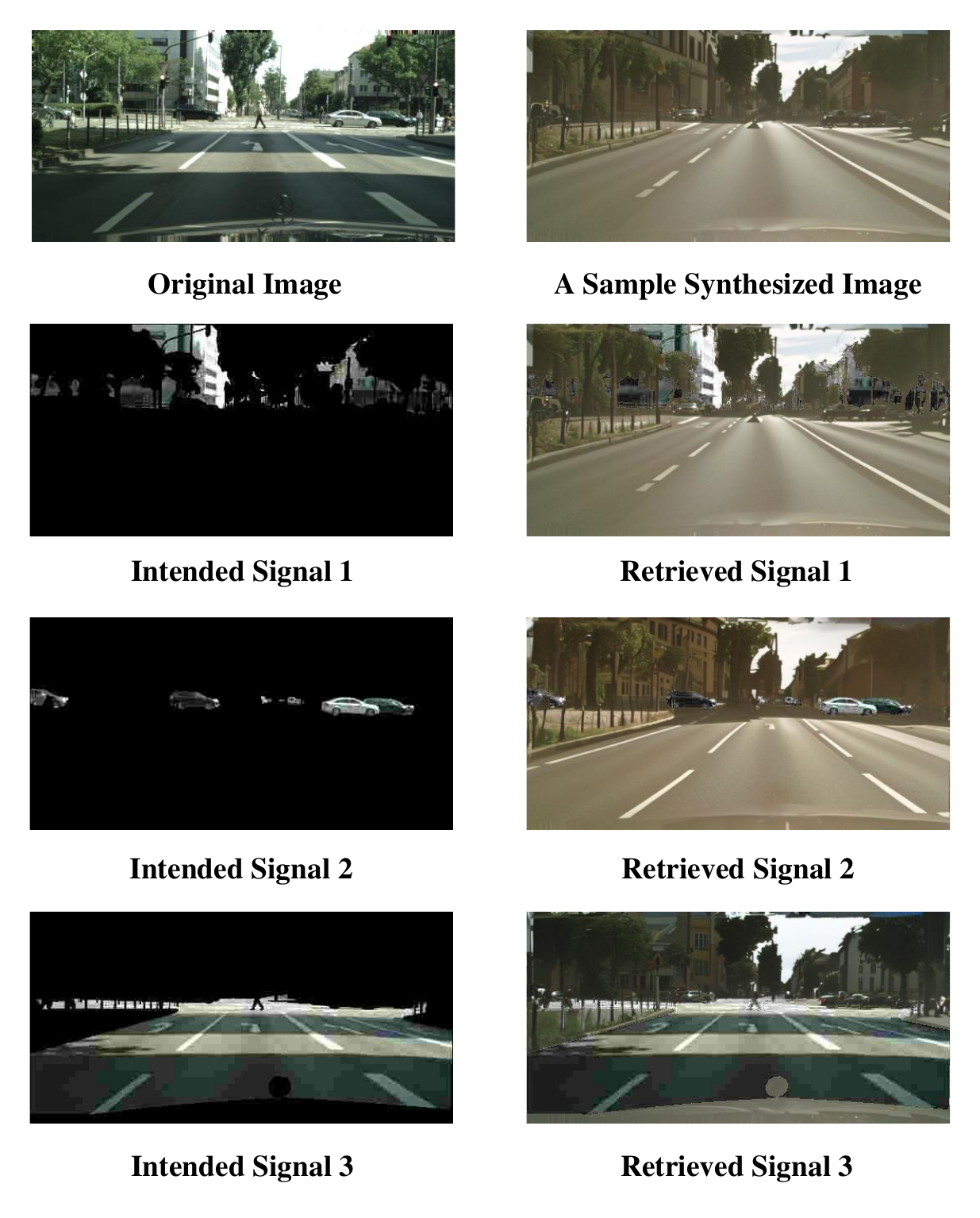}
 \captionsetup{font={footnotesize}, singlelinecheck = off,  name={Fig.},labelsep=period}
\caption{Visual quality of the proposed generative semantic multicasting framework for typical street scene 2, where $(\varepsilon_{kl}^{(r)}, \varepsilon_k^{(s)})=(0.06781, 0.63050)$, $T^*=7.5635$ $\rm{ms}$, $\lambda^* = 6.0154$.}
\label{fig:VQ4}
\end{figure}

\begin{figure}[t]
\centering
\includegraphics[scale=0.43]{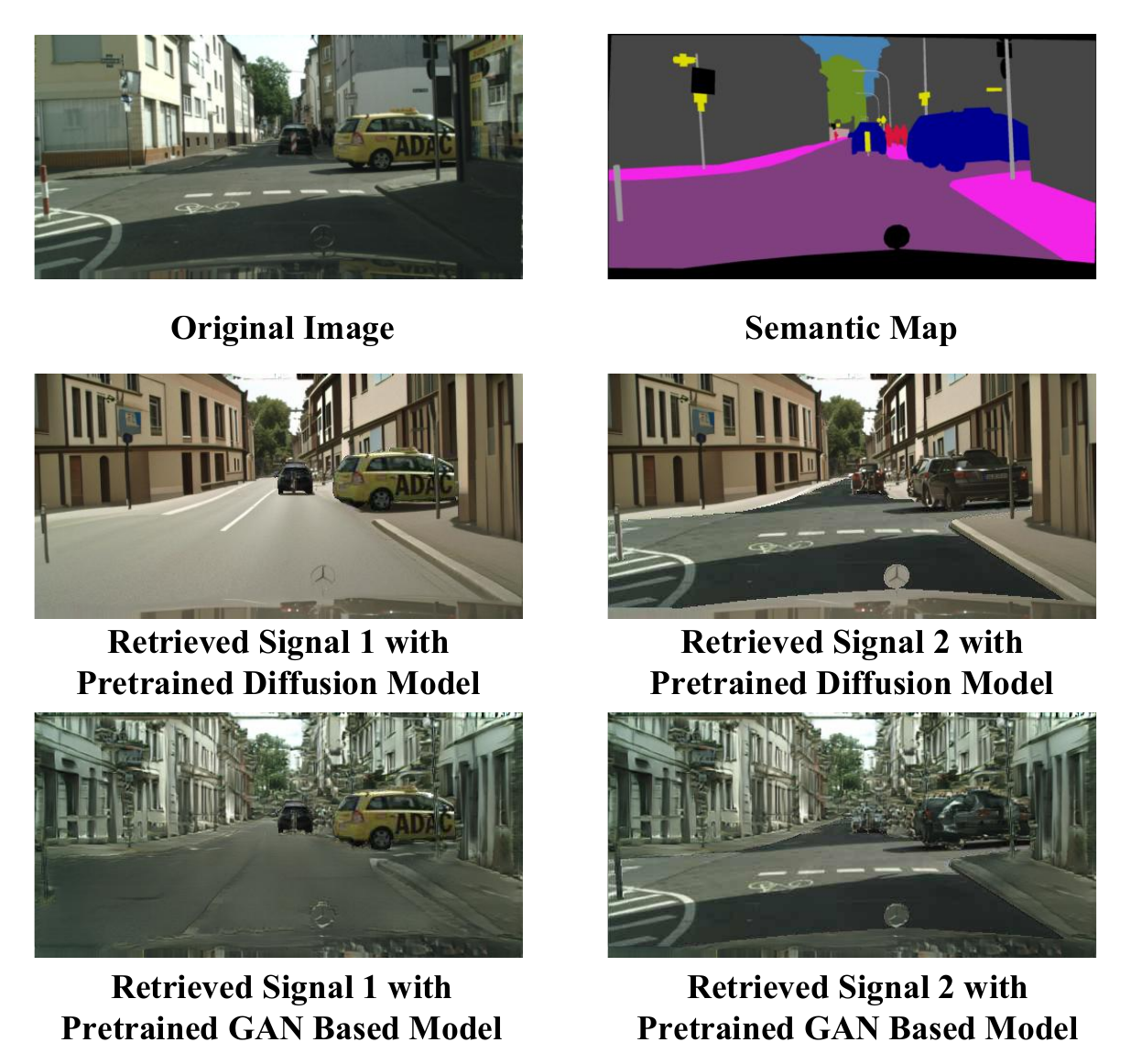}
 \captionsetup{font={footnotesize}, singlelinecheck = off,  name={Fig.},labelsep=period}
\caption{\color{black}Visual quality comparisons for Diffusion vs. GAN-based synthesis.}
\label{fig:VQ5}
\end{figure}

\begin{figure}[t]
\centering
\begin{subfigure}[h]{0.49\columnwidth}
  \centering
  \includegraphics[width=\linewidth]{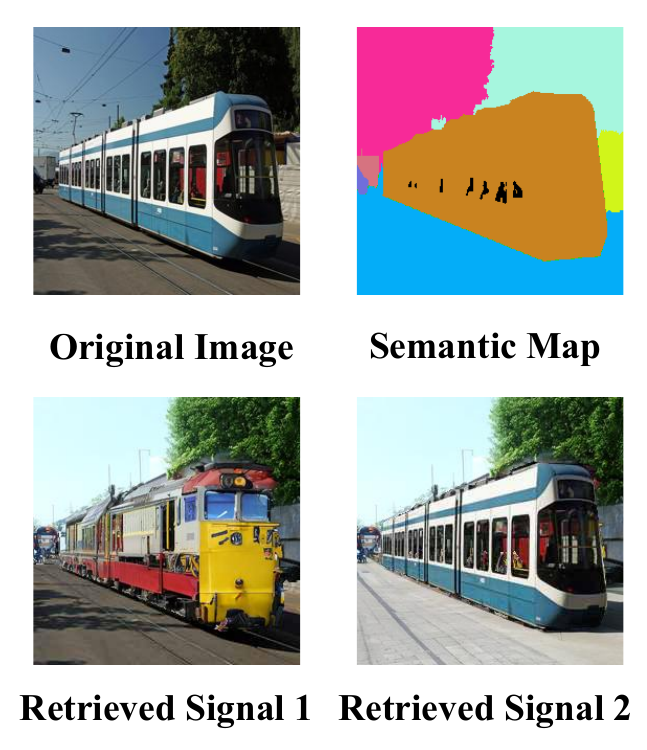}  
  \caption{}
  \label{fig.VQ_COCO(a)}
\end{subfigure}
\begin{subfigure}[h]{0.49\columnwidth}
  \centering
  \includegraphics[width=\linewidth]{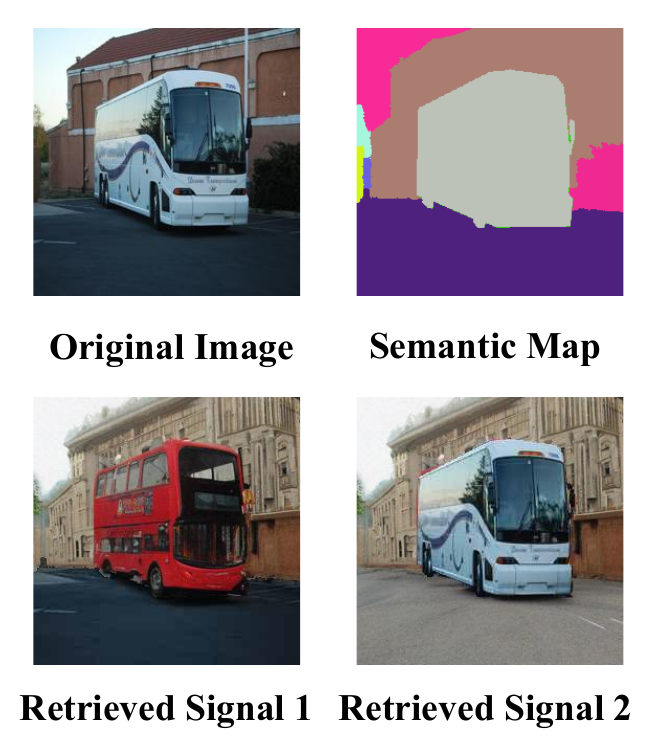}  
  \caption{}
  \label{fig.VQ_COCO(b)}
\end{subfigure}
 \captionsetup{font={footnotesize}, singlelinecheck = off,name={Fig.},labelsep=period}
\caption{\color{black}Visual quality of the proposed generative semantic multicasting framework for typical scene (a) and (b) with COCO-Stuff dataset, where $(\varepsilon_{kl}^{(r)}, \varepsilon_k^{(s)})=(0.02850, 0.58705)$}
\label{fig:VQ_COCO}
\end{figure}

\section{Conclusions}\label{SEC_Conclusion}
In this paper, we developed an adaptive intent-aware framework for generative semantic multicasting, leveraging pre-trained diffusion models. The framework decomposes the source signal into multiple semantic classes based on users' intent, transmitting only the intended classes to each user along with a highly compressed semantic map that allows users to locally synthesize the other classes using a generative model. Furthermore, we designed a communication/computation-aware scheme for per-class adaptation of the communication parameters based on the channel conditions and users' distortion/perception requirements, which significantly improves wireless resource utilization while maintaining the retrieved signal quality. The proposed framework extends beyond images, and can be applied to intent-aware generative multicasting of the emerging multimedia signals, e.g. in XR/VR, metaverse, etc. applications, by adopting the state-of-the-art video/3D generative diffusion models, which we will investigate in future work.

\bibliographystyle{IEEEtran}
\bibliography{Bibliography}

\end{document}